# Fair Isaac Technical Paper

**Subject:**     **Liquid Scorecards**

**From:**      **Bruce Hoadley (BCH – A34  X27051)**

**Date:**      **August 14, 2000**

## Abstract

Traditional scorecards are generalized additive models with bin smoothing. The bins are the attributes of the various characteristics. A traditional scorecard can also be expressed as a sum of the characteristic scores. A characteristic score is a step function of its underlying characteristic. The heights of the steps are just the score weights. If the characteristic is a variable with a lot of values (e.g., Days since last payment), then the step function is somewhat unnatural, because as you cross an attribute boundary the score takes a sudden jump. It is more natural for the score to vary continuously as the characteristic varies continuously.

This paper introduces the liquid scorecard, which is a continuous function of the continuous part of characteristics. The liquid scorecard is still partially discrete, because almost all characteristics have a discrete component, and some characteristics are purely discrete.

The technology used for the liquid scorecard is B-spline theory. This theory provides a completely natural extension of traditional scorecards. In fact, the traditional scorecard is a special case of the liquid scorecard, because the attribute indicator variables form a first order B-spline basis. To get continuity, the liquid scorecard uses a second, third, or fourth order B-spline basis (depending on user preference) as a replacement for some of the attribute indicator variables. The continuous part of the liquid scorecard is then a linear, quadratic, or cubic spline respectively. This means that simple linear regression is also a special case of the liquid scorecard.

A beautiful property of B-splines is that all of score engineering can be handled in essentially the same way as with traditional scorecards. For example, monotonic pattern constraints can be represented as linear inequality constraints involving adjacent liquid score coefficients, which are sort of like score weights – but not exactly.

In addition, the quadratic programming approach to score development, which was introduced in Reference [1], applies to liquid scorecard development in exactly the same way as it applies to traditional scorecards.

The liquid score is not just pretty to look at. It brings real business benefit. In the fraud example featured in Reference [1], the validation divergence of the liquid scorecard is 8.8% larger than for the INFORM11 scorecard. It is 2.9% larger than the INFORM-*NLP* scorecard, and 1.7% larger than the maximum divergence scorecard developed in Reference [1]. And this was done without optimizing the B-spline knots or optimizing the smoothness penalty term. These are subjects of future research.

# Table of Contents

## 1. Introduction

A traditional scorecard can be written as a linear function of attribute indicator variables

$$Score = \sum_{j=1}^{p} S_j X_j ,$$

where the $X_j$'s are the attribute indicator variables and the $S_j$'s are the score weights.

The traditional scorecard is organized into characteristics. Associated with each characteristic are a set of attributes and a corresponding set of indicator variables. For example, consider characteristic 170 (Days since last payment) in the fraud scorecard, S1, featured in Reference [1]. For this characteristic there are 7 attributes with the following score weights:

| Attribute | Score Weight |
|---|---|
| -9,999,999 | 0 |
| -9,999,998 (NO INFORMATION) | 0 |
| 0 - <5 | .306 |
| 5 - <25 | .157 |
| 25 - <35 | -.067 |
| 35 - <300 | -.259 |
| 300 - High | -.888 |

Clearly, the first two attributes are discrete and in-weighted to zero. The rest of the score weights can be plotted as a step function over the range [0,1000] as follows:

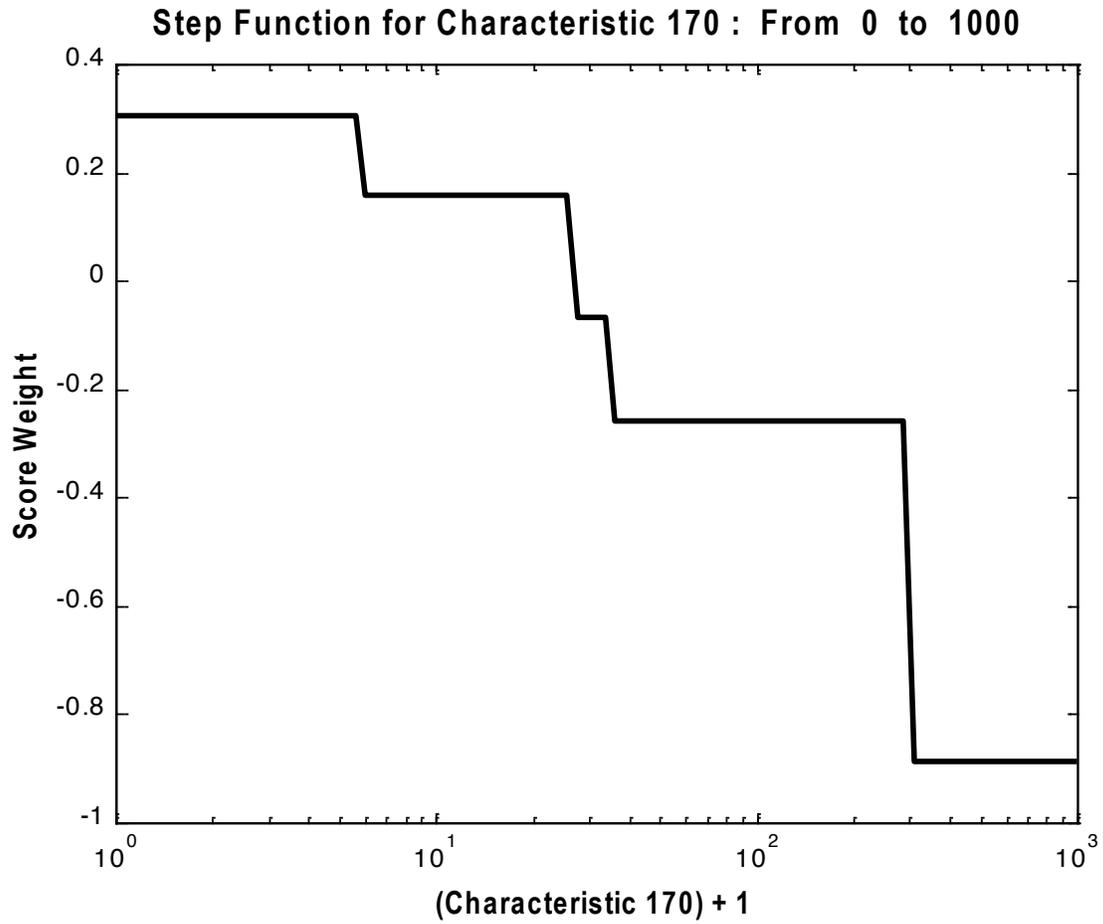

This step function is somewhat unnatural. As characteristic 170 varies from 0 to 1000, it would be more natural for the characteristic score to vary continuously over this range.

A liquid version of this step function would look like

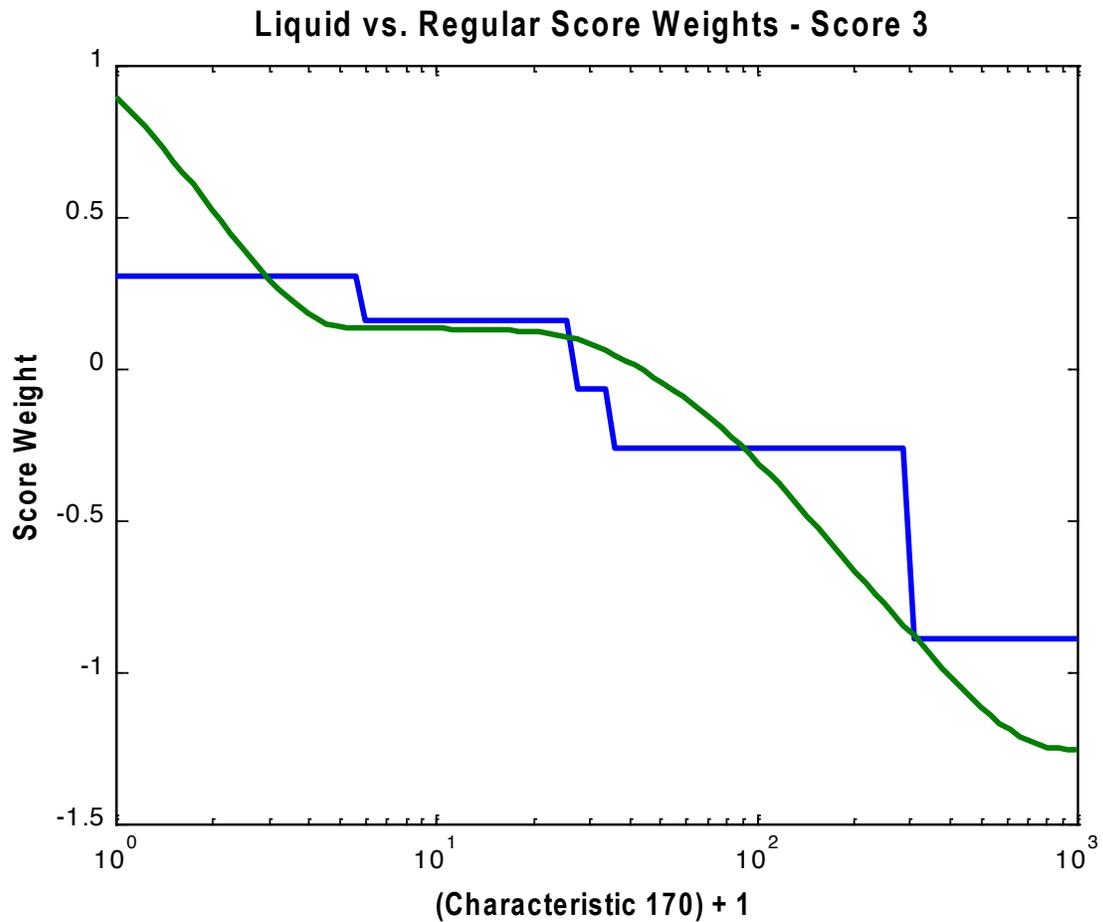

The purpose of this paper is to document the theory, methodology, and MATLAB software for simultaneously fitting these kind of smooth score weight curves, while maintaining a discrete scorecard structure where appropriate, and maintaining all of the traditional score engineering.

The paper is loosely based on the theory of Generalized Additive Models (see Reference [2]). However, the paper is more in the spirit of Reference [3], where B-splines and quadratic programming are used.

## 2. B-Spline Theory

### 2.1 Basis functions

The smooth curve featured in the Introduction is a cubic spline defined over the interval [0,1000]. The cubic spline can be represented as a linear combination of cubic spline basis functions – called B-splines. Associated with the spline basis functions is a set of knots, $\{k(1), k(2), \ldots, k(m)\}$. In the above example, the knots are $\{0, 5, 25, 35, 300, 1000\}$.

The cubic spline basis functions are defined iteratively (see Reference [4]). The first step in this iteration is to define a sequence of numbers, $\{t(1), t(2), \ldots, t(m+6)\}$, which are related to the knots as follows:

$$t(1) = t(2) = t(3) = t(4) = k(1)$$
$$t(i) = k(i-3) \quad \text{for } i = 5, \ldots, m+2$$
$$t(m+3) = t(m+4) = t(m+5) = t(m+6) = k(m).$$

Next I define $4 \times (m+2)$ functions of $x$

$$B(x \mid i, j) \quad i = 1, 2, \ldots, m+2 \quad j = 1, 2, 3, 4.$$

For $j = 1$

$$B(x \mid i, 1) = 1\{t(i) \le x < t(i+1)\} \qquad i = 1, 2, \ldots, m+1$$
$$B(x \mid m+2, 1) = 1\{t(m+2) \le x \le t(m+3)\},$$

where $1_{\{\text{Event}\}}$ is the indicator variable of Event.

For $j = 2, 3, 4$

For $i = 1, 2, ..., m + 2$

$$T_1(x \mid i, j) = \begin{cases} \dfrac{[x - t(i)]}{[t(i + j - 1) - t(i)]} B(x \mid i, j - 1) & \text{if } t(i + j - 1) - t(i) > 0 \\ 0 & \text{otherwise} \end{cases}$$

$$T_2(x \mid i, j) = \begin{cases} \dfrac{[t(i + j) - x]}{[t(i + j) - t(i + 1)]} B(x \mid i + 1, j - 1) & \text{if } t(i + j) - t(i + 1) > 0 \\ 0 & \text{otherwise} \end{cases}$$

$$B(x \mid i, j) = T_1(x \mid i, j) + T_2(x \mid i, j).$$

Some of the functions defined above are vacuous in that they are identically equal to zero. They are

$$B(x \mid 1, 1) = B(x \mid 2, 1) = B(x \mid 3, 1) \equiv 0$$
$$B(x \mid 1, 2) = B(x \mid 2, 2) \equiv 0$$
$$B(x \mid 1, 3) \equiv 0.$$

The functions $B(x \mid 4, 1), B(x \mid 5, 1), ..., B(x \mid m + 2, 1)$ form a first order B-spline basis with knots, $\{k(1), ..., k(m)\}$. This means that every step function with these knots can be expressed as a linear combination of $B(x \mid 4, 1), B(x \mid 5, 1), ..., B(x \mid m + 2, 1)$. These $m - 1$ basis functions are just the attribute indicator variables for the attributes defined by the knots.

The functions $B(x \mid 3, 2), B(x \mid 4, 2), ..., B(x \mid m + 2, 2)$ form a second order B-spline basis with knots, $\{k(1), ..., k(m)\}$. This means that every piecewise linear function with these knots can be expressed as a linear combination of $B(x \mid 3, 2), B(x \mid 4, 2), ..., B(x \mid m + 2, 2)$.

The functions $B(x \mid 2, 3), B(x \mid 3, 3), ..., B(x \mid m + 2, 3)$ form a third order B-spline basis with knots, $\{k(1), ..., k(m)\}$. This means that every quadratic spline with these knots can be expressed as a linear combination of $B(x \mid 2, 3), B(x \mid 3, 3), ..., B(x \mid m + 2, 3)$. These quadratic splines have matching first derivatives at the internal knots, $\{k(2), ..., k(m - 1)\}$.

The functions $B(x\,|\,1,4), B(x\,|\,2,4),..., B(x\,|\,m+2,4)$ form a fourth order B-spline basis with knots, $\{k(1),...,k(m)\}$. This means that every cubic spline with these knots can be expressed as a linear combination of $B(x\,|\,1,4), B(x\,|\,2,4),..., B(x\,|\,m+2,4)$. These cubic splines have matching first and second derivatives at the internal knots, $\{k(2),...,k(m-1)\}$.

## 2.2 Plots of the basis functions.

In order to visualize the B-spline basis functions, I am going to plot some of them in this section. For the plots, I will use the simple set of knots, {0,1,2,3,4,5}. In this example, $m=6$, so there are 5 first order basis functions, 6 second order basis functions, 7 third order basis functions, and 8 fourth order basis functions.

The first order basis functions are just the indicator functions for the attributes: [0,1), [1,2), [2,3), [3,4), and [4,5). The plot of these indicator variables is uninteresting, so I do not plot them.

The second order basis functions are composed of linear splines and look like half or full tents. The first one has the range [0,1), and has values going from 1 to zero. The second one has the range [0,2), and is a tent with the top value at x=1. The third one has the range [1,3), and is a tent with the top value at x=2. The fourth one has the range [2,4), and is a tent with the top value at x=3. The fifth one has the range [3,5), and is a tent with the top value at x=4. The last one has the range [4,5] and has values going from zero to one.

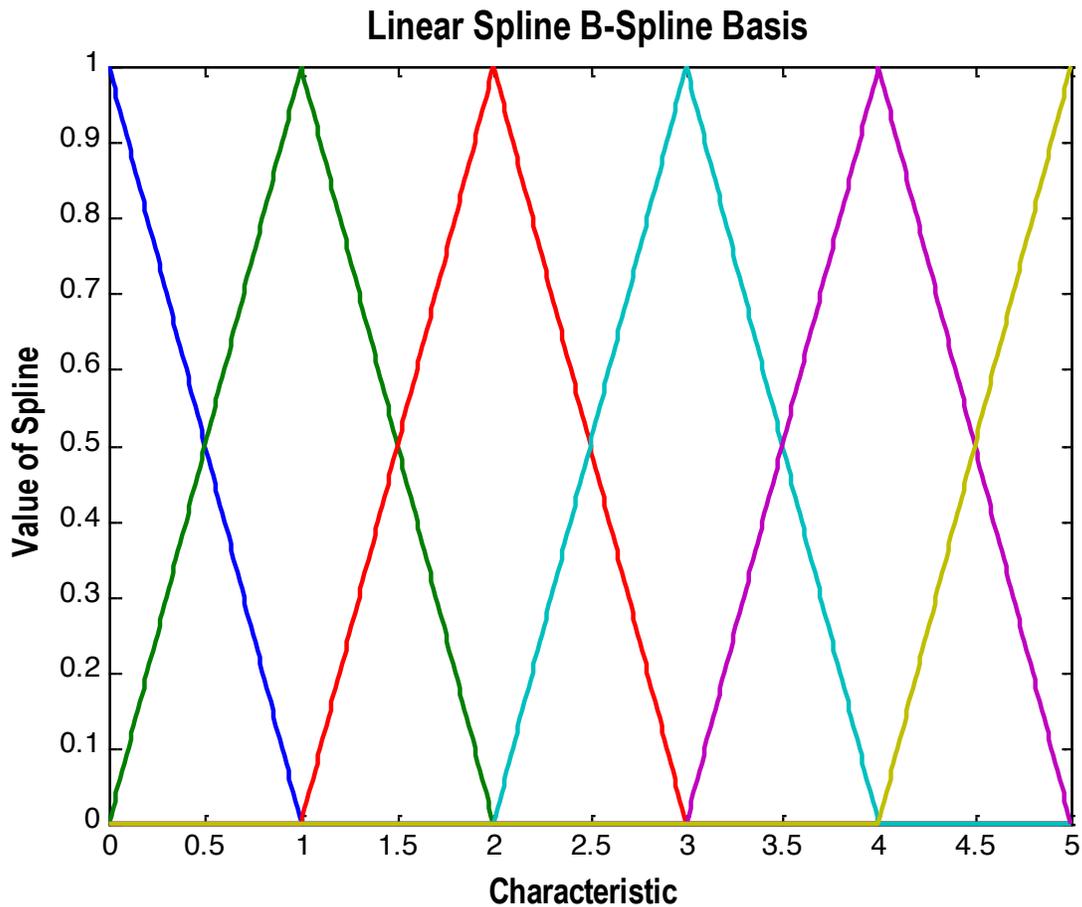

The third order basis functions are composed of quadratic splines and look like half or full bell curves. The first one has the range [0,1), and has values going from 1 to zero. The second one has the range [0,2), and is a bell curve. The third one has the range [0,3), and is a bell curve. The fourth one has the range [1,4), and is a bell curve. The fifth one has the range [2,5), and is a bell curve. The sixth one has the range [3,5), and is a bell curve. The last one has the range [4,5] and has values going from zero to one.

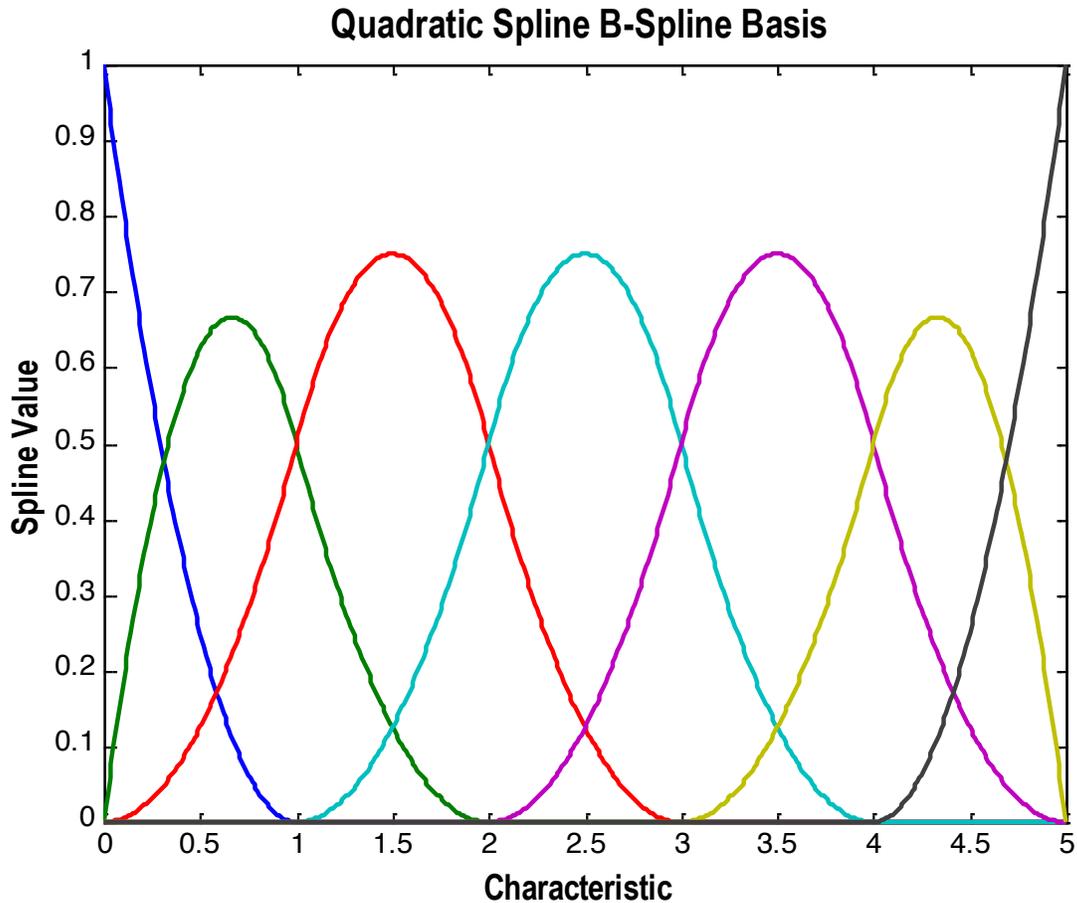

The fourth order basis functions are composed of cubic splines and look like half or full bell curves. The first one has the range [0,1), and has values going from 1 to zero. The second one has the range [0,2), and is a bell curve. The third one has the range [0,3), and is a bell curve. The fourth one has the range [0,4), and is a bell curve. The fifth one has the range [1,5), and is a bell curve. The sixth one has the range [2,5), and is a bell curve. The seventh one has the range [3,5), and is a bell curve. The last one has the range [4,5] and has values going from zero to one.

**Cubic Spline B-Spline Basis**

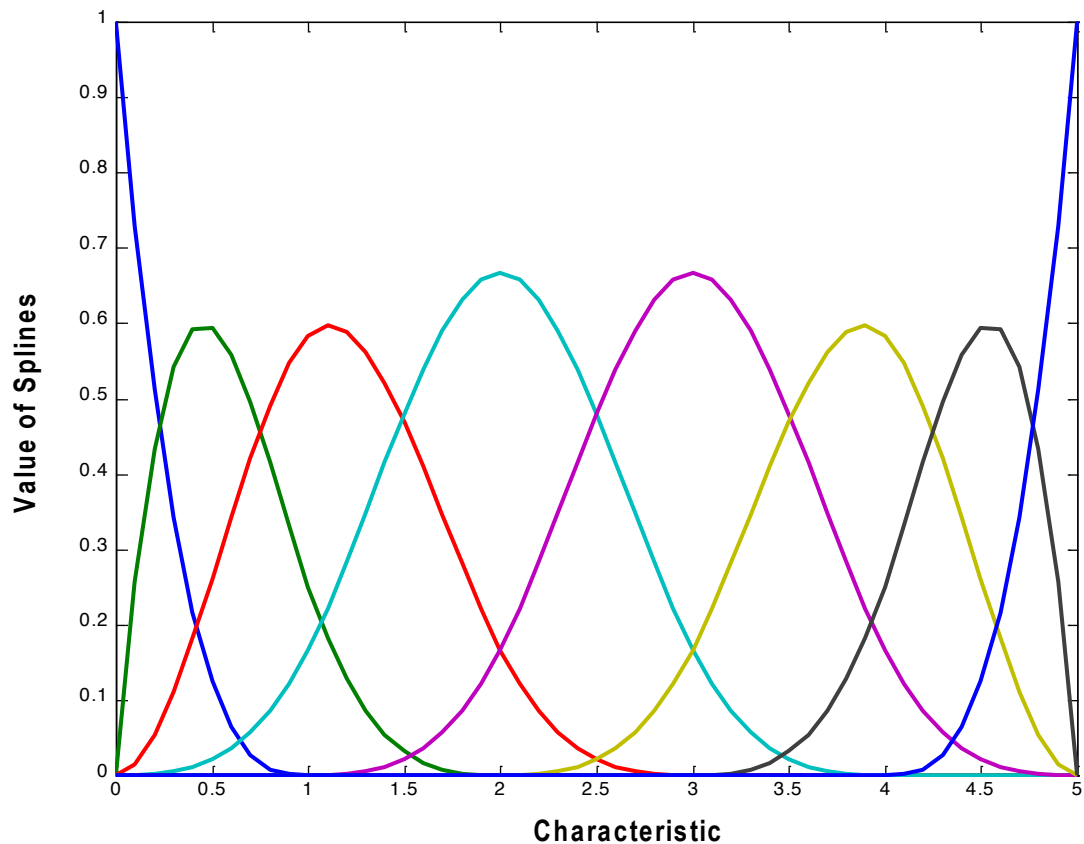

## 2.3 MATLAB code for plotting the B-spline basis functions

```
» x=0:.01:5;
» x=x';
» knots=0:5;
» B=bspline(x,knots);
» plot(x,B(:,11),x,B(:,12),x,B(:,13), ...
      x,B(:,14),x,B(:,15),x,B(:,16));
» plot(x,B(:,18),x,B(:,19),x,B(:,20), ...
      x,B(:,21),x,B(:,22),x,B(:,23),x,B(:,24));
» plot(x,B(:,25),x,B(:,26),x,B(:,27), ...
       x,B(:,28),x,B(:,29),x,B(:,30), ...
       x,B(:,31),x,B(:,32));
```

In the MATLAB code above, I use the MATLAB function, bspline. The MATLAB code, for all of the MATLAB functions used in this paper, are in Appendix 1.

## 2.4 Properties of the B-spline basis functions

I defined above four sets of basis functions, which can be used to represent functions defined on the interval, $[k(1), k(m)]$. Mathematically, I denote the first order basis functions by

$$b_{14}(x), b_{15}(x), ..., b_{1,m+2}(x).$$

I denote the second order basis functions by

$$b_{23}(x), b_{24}(x), ..., b_{2,m+2}(x).$$

I denote the third order basis functions by

$$b_{32}(x), b_{33}(x), ..., b_{3,m+2}(x).$$

I denote the fourth order basis functions by

$$b_{41}(x), b_{42}(x), ..., b_{4,m+2}(x).$$

Any step function defined on the knots $\{k(1), ..., k(m)\}$ can be written as

$$c_1(x) = \sum_{j=4}^{m+2} \alpha_{1j} b_{1j}(x).$$

Any piecewise linear function defined on the knots $\{k(1), \ldots, k(m)\}$ can be written as

$$c_2(x) = \sum_{j=3}^{m+2} \alpha_{2j} b_{2j}(x).$$

Any quadratic spline function defined on the knots $\{k(1), \ldots, k(m)\}$ can be written as

$$c_3(x) = \sum_{j=2}^{m+2} \alpha_{3j} b_{3j}(x).$$

Any cubic spline defined on the knots $\{k(1), \ldots, k(m)\}$ can be written as

$$c_4(x) = \sum_{j=1}^{m+2} \alpha_{4j} b_{4j}(x).$$

There is a very close relationship between the nature of the spline functions, $\{c_i(x)\}$ and the coefficients of the basis functions, $\{\alpha_{ij}\}$. I state these as a set of properties.

1. The value of the function at the left knot is the first coefficient; i.e.,

$$c_1(k(1)) = \alpha_{14}$$
$$c_2(k(1)) = \alpha_{23}$$
$$c_3(k(1)) = \alpha_{32}$$
$$c_4(k(1)) = \alpha_{41}.$$

2. The value of the function at the right knot is the last coefficient; i.e.,

$$c_i(k(m)) = \alpha_{i,m+2}.$$

3. For a fixed value of $x \in [k(1), k(m)]$, the sum over any set of B-spline basis functions is one; i.e.,

$$\sum_{j=4}^{m+2} b_{1j}(x) = 1$$

$$\sum_{j=3}^{m+2} b_{2j}(x) = 1$$

$$\sum_{j=2}^{m+2} b_{3j}(x) = 1$$

$$\sum_{j=1}^{m+2} b_{4j}(x) = 1.$$

So any set of B-spline basis functions behaves like generalized indicator variables. Of course, the first order splines are indicator variables. The second, third, and fourth order B-splines are like fuzzy indicator variables.

4. The spline, $c_i(x)$, is monotonically increasing if and only if its associated coefficients are monotonically increasing; i.e., $\alpha_{ij} \le \alpha_{i,j+1}$, for all $j$.

5. The spline, $c_i(x)$, is monotonically decreasing if and only if its associated coefficients are monotonically decreasing; i.e., $\alpha_{ij} \ge \alpha_{i,j+1}$, for all $j$.

6. The number of changes in direction for the spline, $c_i(x)$, is equal to the number of changes in direction for the associated coefficients, $\{\alpha_{ij}\}$. For example, if $\alpha_{41} \ge \alpha_{42} \ge \alpha_{43}$ and $\alpha_{43} \le \alpha_{44} \le \ldots \le \alpha_{4,m+2}$, then the spline function, $c_4(x)$, will be at first monotonically decreasing and then become monotonically increasing.

These properties are described in References [3] and [4].

To illustrate these properties, consider again the example where the knots are {0, 1, 2, 3, 4, 5}. And consider the function

$$c_4(x) = (1)b_{41}(x) + (0)b_{42}(x) + (-1)b_{43}(x) + (0)b_{44}(x)$$
$$+ (.5)b_{45}(x) + (1)b_{46}(x) + (1.5)b_{47}(x) + (2)b_{48}(x).$$

A plot of this cubic spline is

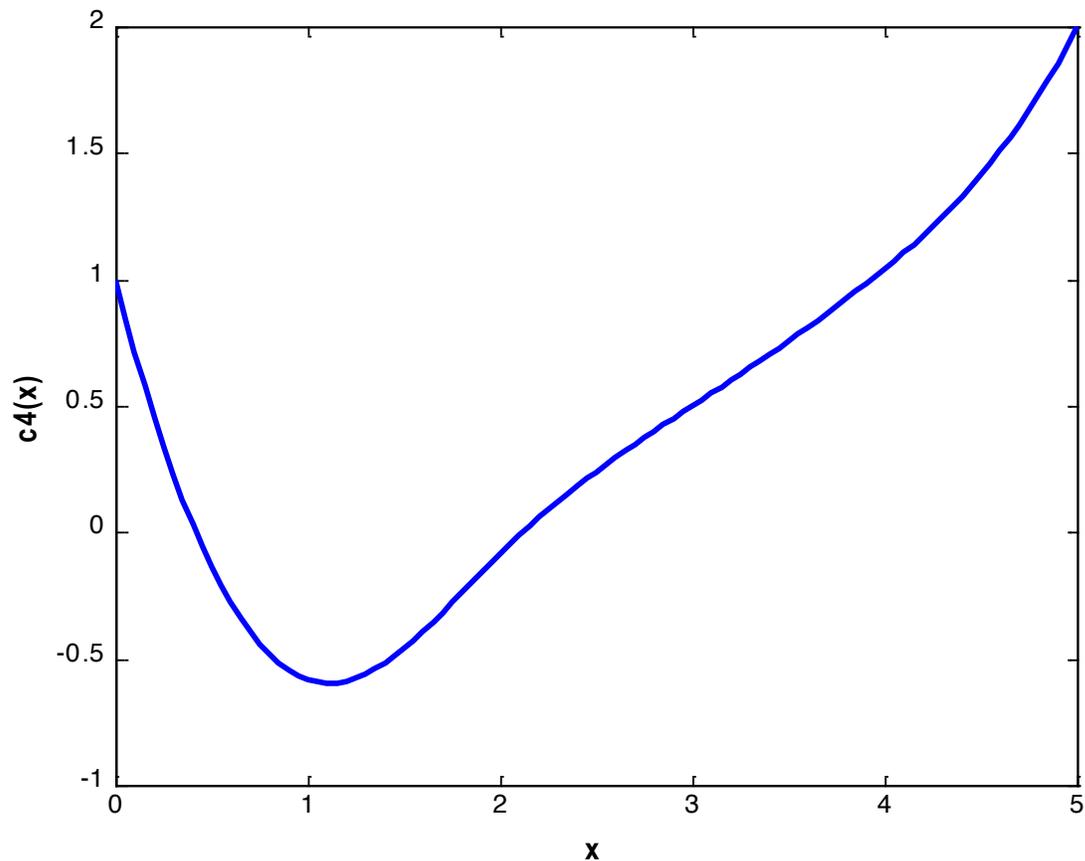

As advertised, the value of the function is 1 at x=0, and the value of the function is 2 at x=5, and the function starts out decreasing and ends up increasing.

These properties make it very easy to control the shape of the function by controlling the patterns in the coefficients.

Hence, the B-splines are ideal for score engineering. Traditional scorecards are based on step functions; i.e., first order B-splines. Traditional scorecard engineering is handled by linear equality and linear inequality constraints on the score weights; i.e., the $\alpha_{1j}$'s. For scores based on second, third or fourth order B-splines, score engineering can also be handled by linear equality and linear inequality constraints on the score weights; i.e., the $\alpha_{2j}$'s, $\alpha_{3j}$'s, and $\alpha_{4j}$'s respectively.

This means that the quadratic programming approach to score development, which was described in Reference [1], applies to score components based on the B-spline basis functions.

## 2.5 Simple linear regression

If you want the score formula to be a simple linear function of a characteristic over some range, then you use the simple knots, $\{k(1), k(2)\}$, and use a second order spline basis. In this case, $m = 2$, and there are two linear functions in the second order B-spline basis.

# 3. Quadratic Programming Formulation of Liquid Scorecard Development

## 3.1 Liquid Score formula

The liquid score formula is still of the form

$$Score = \sum_{j=1}^{p} S_j X_j \, ,$$

where the $X_j's$ are either attribute indicator variables or B-spline basis functions. In this paper, I use the cubic spline basis functions for the liquid part of the scorecard. These basis functions are numerical predictors, just like the attribute indicator variables. So the quadratic programming theory developed in Reference [1] applies.

For traditional scorecards, the $S_j's$ are score weights. For liquid scorecards, I will call them liquid score coefficients. Of course, the $S_j's$ associated with the attribute indicator variables in the liquid scorecard are also score weights.

The score can be written in matrix notation as

$$Score = \boldsymbol{S'} * \boldsymbol{X} \, ,$$

where

$$\boldsymbol{S'} = \left(S_1, ..., S_p\right)$$
$$\boldsymbol{X'} = \left(X_1, ..., X_p\right).$$

I use bold to indicate a matrix or vector and use $*$ to indicate ordinary matrix multiplication.

## 3.2 Classic quadratic program

In Reference [1], Section 2.1, I showed that the problem of maximizing divergence subject to score engineering (with no non-zero in-weighting) can be formulated as the following quadratic program:

$$\text{Find } S \text{ to}$$
$$\text{Minimize} \quad S' * C * S$$
$$\text{Subject to :}$$
$$Ac * S = 0$$
$$d' * S = \delta$$
$$Ap * S \leq 0 ,$$

where

$$C = \frac{Cov[X \mid G] + Cov[X \mid B]}{2}$$
$$d = E[X \mid G] - E[X \mid B].$$

The matrices, $Ac$ and $Ap$, define the linear equality and linear inequality constraints, respectively.

So the matrices in the general form of the MATLAB quadratic program (see Reference [1], Section 2.3) are

$$H = 2C$$
$$f = 0$$
$$Aeq = \begin{bmatrix} d' \\ Ac \end{bmatrix}$$
$$beq = \begin{bmatrix} \delta \\ 0 \end{bmatrix}$$
$$A = Ap$$
$$b = 0$$
$$l = -\infty$$
$$u = +\infty .$$

To apply quadratic programming to the liquid scorecard, I have to compute theses matrices. In the MATLAB code, they are called

```
C3, d3, H3, f3, Aeq3, beq3, A3, b3, l3, u3
```

## 4. Development of a Liquid Scorecard

To illustrate the development of a liquid scorecard, I will use the same fraud example featured in Reference [1]. The traditional fraud scorecard structure of characteristics and attributes is shown in Appendix 2 of Reference [1].

For this application I use cubic B-spline basis functions for each liquid characteristics. However, I could have used linear or quadratic B-spline basis functions.

### 4.1 Liquid characteristics

It is clear that some of the characteristics in the fraud scorecard are purely discrete and have no liquid component. An exploratory data analysis of the characteristics revealed that Characteristics 170, 191, 193, 211, 260, 320, 380, 330, 710, 960, 961, 962, and 965 have liquid components.

### 4.2 Knots for the liquid characteristics

In my first attempt at a liquid scorecard, I used only two internal knots for each liquid characteristic. These knots were placed at roughly the 33rd and 67th percentile of the characteristic distributions. This did not work. The liquid scorecard had a development divergence of 1.653, which is lower than the development divergence of the traditional scorecard, which is 1.753.

It is easy to see why this did not work. A plot of the traditional characteristic score vs. the liquid characteristic score for Characteristic 191 is

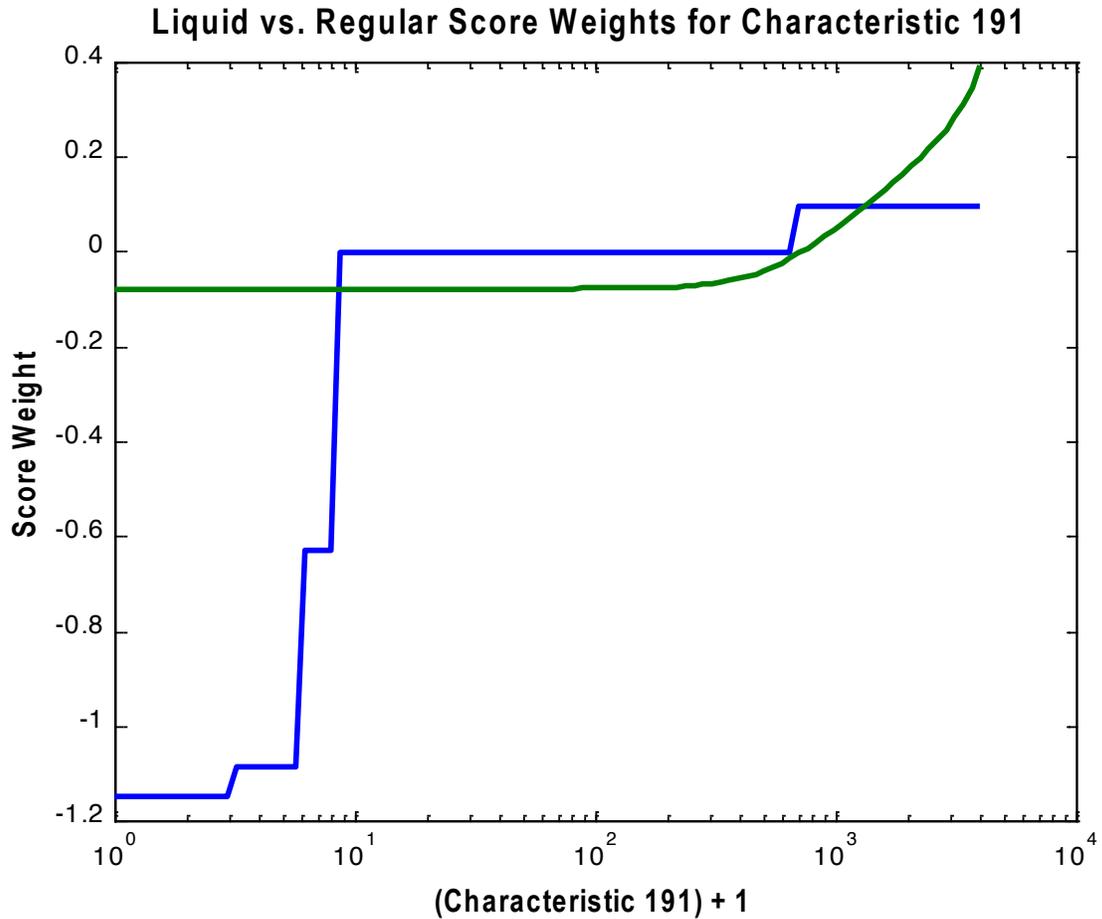

Here the liquid knots were {0, 265, 725, 4000}. The rapid change of the characteristic score between the characteristic values of 2 and 10 is completely missed by the liquid score. This convinced me to use the knots of the traditional scorecard as the liquid knots. The analyst, who had established those knots, had done a careful analysis. So I might as well use them.

The following Table lists the liquid characteristics along with the knot structures for the liquid components of these characteristics.

## Liquid Characteristics and their Knots

| Liquid Characteristic | Knots |
|:---:|:---:|
| 170 | {0, 5, 25, 35, 300, 1000} |
| 191 | {0, 2, 5, 7, 650, 4000} |
| 193 | {0, 1, 2, 3, 18, 1000} |
| 211 | {0, 1, 7, 35, 80, 200, 400, 800, 1300, 1700, 10000} |
| 260 | {0, 101, 210, 305, 565, 700, 1500} |
| 320 | {0, 590, 2055, 8405, 16960, 20000, 30000, 40375, 70000, 300000} |
| 380 | {0, 635, 1210, 1915, 5000, 10000} |
| 330 | {1, 250, 4000} |
| 710 | {1, 360, 675, 2435, 14000} |
| 960 | {-2950, -1700, -800, -450, 1425} |
| 961 | {-2950, -1700, -800, 550, 1425} |
| 962 | {-2950, -1500, -1100, -850, -550, -400, -300, 1, 200, 1425} |
| 965 | {-2950, -950, -750, -550, -400, -300, -200, -100, 80, 1425} |

For characteristics 170 through 710, the maximum knot was chosen to be at about the 99th percentile of the distribution for the characteristic. For these liquid characteristics, all values above this upper bound were collapsed to the upper bound. For characteristics 960 through 965, the minimum and maximum knots were chosen approximately to be the minimum and maximum of the characteristic, respectively.

## 4.3 Exploratory data analysis

To arrive at the structure above, I did some exploratory data analysis. Here I record some of the simple MATLAB commands that do the exploratory data analysis.

I start with the data set described in Section 2.5 of Reference [1]. This is the $14000 \times 206$ MATLAB matrix called Frdata.

Consider Characteristic 170, which is the 6th column of the Frdata matrix. The MATLAB command for putting Characteristic 170 into a column vector is

```
c170=Frdata(:,6);
```

The two purely discrete values of Characteristic 170 are –9,999,999 and –9,999,998. To compute the vector of values of c170, which are more than –9999998 (or $\geq 0$), use the MATLAB command

```
c170c=c170(c170>-9999998);
```

This is the liquid part of Characteristic 170.

The MATLAB commands for computing the percentiles of `c170c` are

```
p=(0:1:100);
pc170c=prctile(c170c,p);
```

These percentiles were used to pick 1,000 as the upper truncation point.

To compute the boxplot of c170c for values of c170c, which are less than 1000, use the MATLAB command

```
boxplot(c170c(c170c<1000))
```

## 4.4 Liquid scorecard structure

The structure of the liquid scorecard is shown in Appendix 2. The rows of this liquid scorecard are divided into three sections. The first 68 rows contains the purely discrete characteristics. They are Characteristics 471, 503, 533, 635, 665, 830, 835, 840, 843, 860, 870, and 950. There are 12 in all. Rows 69 through 94 show the discrete parts of the liquid characteristics. Rows 95 through 210 show the liquid part of the liquid characteristics. There are 13 liquid characteristics. The reason for organizing it this way is to facilitate experimentation with the knots of the liquid components.

The second column shows the attribute structure for the discrete parts of the scorecard and the knot structure for the liquid parts of the scorecard. Consider the liquid part of Characteristic 170. The rows are labeled #, 0-<5,…,300-High, 1000, #. So there are three more rows than there are traditional attributes. This is because the number of liquid score coefficients (i.e., the number of cubic spline basis functions) is three more than the number of traditional attributes. The knots for the cubic splines are 0, 5, 25, 35, 300, and 1000. The rows labeled by the # symbol will contain the first and last liquid score coefficients. Recall that the first liquid score coefficient is the value of the score weight function at the left end knot, and the last liquid score coefficient is the value of the score weight function at the right end knot. So these liquid score coefficients are very interpretable.

The third column shows the liquid score coefficient number. There are 210 liquid score coefficients.

The fourth column shows the constraints associated with score engineering. For the discrete part of the scorecard, these are the same as for the traditional scorecard. For the liquid part of the scorecard, they are deduced from the pattern constraints of the traditional scorecard.

The fifth column shows the column numbers in the Frdata matrix for the associated attribute indicator variables. Note that this column is filled only for the discrete part of the scorecard. This is because, for the liquid part, I will use the cubic B-spline basis functions rather than the attribute indicator variables.

The sixth column shows the S1 score weights developed in Appendix 2 of Reference [1], for the discrete part of the scorecard. These are shown for comparison purposes. These score weights should be very similar to the score coefficients for the discrete part of the liquid scorecard.

## 4.5 Defining the design matrix for the liquid scorecard

### The discrete part of the liquid scorecard

As we can see from Appendix 2, the first 94 variables in the liquid scorecard are the indicator variables for the discrete attributes. The MATLAB code for putting these 94 variables into a design matrix is

```
» X=zeros(14000,94);
» X(:,1:25)=Frdata(:,100:124);
» X(:,26:68)=Frdata(:,133:175);
» X(:,69)=Frdata(:,36);
» X(:,70:71)=Frdata(:,42:43);
» X(:,72:73)=Frdata(:,49:50);
» X(:,74:76)=Frdata(:,56:58);
» X(:,77:78)=Frdata(:,69:70);
» X(:,79:80)=Frdata(:,77:78);
» X(:,81)=Frdata(:,88);
» X(:,82)=Frdata(:,93);
» X(:,83)=Frdata(:,99);
» X(:,84)=Frdata(:,89);
» X(:,85)=Frdata(:,92);
» X(:,86:88)=Frdata(:,125:127);
» X(:,89)=Frdata(:,132);
» X(:,90:91)=Frdata(:,180:181);
» X(:,92)=Frdata(:,186);
» X(:,93)=Frdata(:,196);
» X(:,94)=Frdata(:,206);
```

### Capturing the liquid characteristics

The 13 liquid characteristics are specified in the last part of Appendix 2. The MATLAB code for capturing and truncating (where necessary) these liquid characteristics is

```
» LC3=zeros(14000,13);
» LC3(:,1)=min(Frdata(:,6),1000);
» LC3(:,2)=min(Frdata(:,7),4000);
» LC3(:,3)=min(Frdata(:,8),1000);
```

```
»   LC3(:,4)=min(Frdata(:,9),1000);
»   LC3(:,5)=min(Frdata(:,10),10000);
»   LC3(:,6)=min(Frdata(:,11),15000);
»   LC3(:,7)=min(Frdata(:,13),10000);
»   LC3(:,8)=min(Frdata(:,12),4000);
»   LC3(:,9)=min(Frdata(:,19),14000);
»   LC3(:,10)=Frdata(:,27);
»   LC3(:,11)=Frdata(:,28);
»   LC3(:,12)=Frdata(:,29);
»   LC3(:,13)=Frdata(:,30);
```

## Definition of the knots for the liquid characteristics

Here I am going to use the same knots that were used to define the attributes in the traditional scorecard. The 13 sets of knots are captured in a MATLAB cell array

```
attk ={[0 5 25 35 300 1000] ...
       [0 2 5 7 650 4000] ...
       [0 1 2 3 18 1000] ...
       [0 1 7 35 80 200 400 800 1300 1700 10000] ...
       [0 101 210 305 565 700 1500] ...
       [0 590 2055 8405 16960 20000 30000 40375 70000 300000]
...
       [0 635 1210 1915 5000 10000] ...
       [1 250 4000] ...
       [1 360 675 2435 14000] ...
       [-2950 -1700 -800 -450 1425] ...
       [-2950 -1700 -800 550 1425] ...
       [-2950 -1500 -1100 -850 -550 -400 -300 1 200 1425] ...
       [-2950 -950 -750 -550 -400 -300 -200 -100 80 1425]};
```

## Define liquid prediction variables

The following MATLAB code constructs the cubic B-spline basis functions for each liquid characteristic and appends them to the design matrix. The code for the MATLAB function bdesign and all other MATLAB functions (developed by me) are in Appendix 1.

```
for i=1:13
    BX=bdesign(LC3(:,i),attk{i});
    X=[X BX];
End
```

The design matrix for the liquid scorecard is now complete.

## 4.6 Development and validation design matrices and performance variables

The MATLAB code is

```
» y=Frdata(:, 32);
» sn=Frdata(:, 3);
» yv=y(((sn==1)|(sn==4)|(sn==8)));
» yd=y(~((sn==1)|(sn==4)|(sn==8)));
» Xd=X(~((sn==1)|(sn==4)|(sn==8)),:);
» Xv=X(((sn==1)|(sn==4)|(sn==8)),:);
```

## 4.7 Compute C, d, e and H matrices required for quadratic program

In Reference [1] I went into a lot of detail on how to set up quadratic programs for solving score development problems. There I described the matrices, called C,d,e, and H, which are required to formulate score development quadratic programs. The MATLAB code is

```
[C3 d3 e3]=divstats(Xd,yd);
H3=Hmatrix(C3,0);
```

See Appendix 1 for details on the MATLAB functions, divstats and Hmatrix.

## 4.8 Computing the Aeq matrix for the linear equality constraints

The Aeq matrix is used in the quadratic program to define the left-hand side of the linear equality constraints.

## <u>In-weighting</u>

The first group of rows in the Aeq matrix is used to model the in-weighting. The zero in-weighted liquid score coefficients are defined by

```
»  iw=[1 4 7 14 21 25 29 35 40 45 48 52 68 ...
       69:72 74 77 79 81:83 85:87 89 90 92:94];
```

The MATLAB function (see Appendix 1)

```
»  Aeqiw3=Aeqiw(iw,210);
```

creates the in-weighting rows.

## Cross restrictions

Each row of the cross3 matrix shows the liquid score coefficients for a cross restriction.

```
»  cross3=[5 26;26 46;46 49]
```

The MATLAB function (see Appendix 1)

```
»  Aeqcross3=Aeqcross(cross3,210);
```

creates the cross restriction rows of Aeq.

## Centering

The liquid score coefficients associated with each characteristic are given in the following MATLAB cell array.

```
»  charx={[1:4] [5:7] [8:14] [15:21] [22:25] ...
          [26:29] [30:35] [36:40] [41:45] [46:48] ...
          [49:52] [53:68] [69 70 95:102] ...
          [71 72 103:110] [73 74 111:118] [75:77 119:131]
       ...
          [78 79 132:140] [80 81 141:152] [82 83 153:160]
       ...
          [84 85 161:165] [86:89 166:172] [90 173:179] ...
          [91 92 180:186] [93 187:198] [94 199:210] };
```

The MATLAB function (see Appendix 1)

```
       Aeqcen3 = Aeqcenter(charx,e3);
```

creates the centering rows of Aeq.

## Final computation of Aeq3

Now I put all these rows together and add a first row, which models the linear divergence constraint (see Reference [1]). The MATLAB code is

```
       Aeq3=[d3;Aeqiw3;Aeqcross3;Aeqcen3];
```

## 4.9 Computing the A matrix for the linear inequality (pattern) constraints

The liquid score coefficients associated with the left-hand side of the pattern constraints are given by

```
» left3=[2 5 5 9:12 16:19 22 23 26 27 30:33 ...
  36:38 46 49 50 53:66 80 84 88 ...
  95:101 103:109 111:117 119:121 ...
  124:130 132:139 141:151 153:159 ...
  161:164 166:171 173:178 180:185 ...
  187:197 199:209];
```

The length of this vector is 143, so there are 143 pattern constraints.

In most cases, the liquid score coefficient associated with the right-hand side of a pattern constraint is one more than for the right-hand side. Hence

```
» right3=left3+1;
```

However, there are exceptions, which are

```
» right3(3)=7;
```

```
» right3(40)=141;
```

```
» right3(41)=161;
```

```
» right3(42)=166;
```

A pattern constraint can be a less than relationship or a greater than relationship. The vector, called ilessthan3, flags those pattern constraints, which are the less than type.

```
» ilessthan3=zeros(1,143);
```

```
» ilessthan3(1)=1;
```

```
» ilessthan3(50:63)=1;
```

```
» ilessthan3(67:73)=1;
```

```
» ilessthan3(110:143)=1;
```

The MATLAB function (see Appendix 1)

```
A3=patterns(left3,right3,ilessthan3,210);
```

creates the A matrix for the pattern constraints.

In my first attempt at developing this particular liquid scorecard, the pattern constraints for the liquid part of Characteristic 211 were

$$S_{119} \geq ... \geq S_{122}$$
$$S_{126} \leq ... \leq S_{131} \,.$$

I knew that the score weights should be decreasing at first and then increasing. But I did not know where the bottom was. I did not pattern constrain all of the liquid score coefficients, because I wanted to leave some freedom for the data to find the bottom. The result was

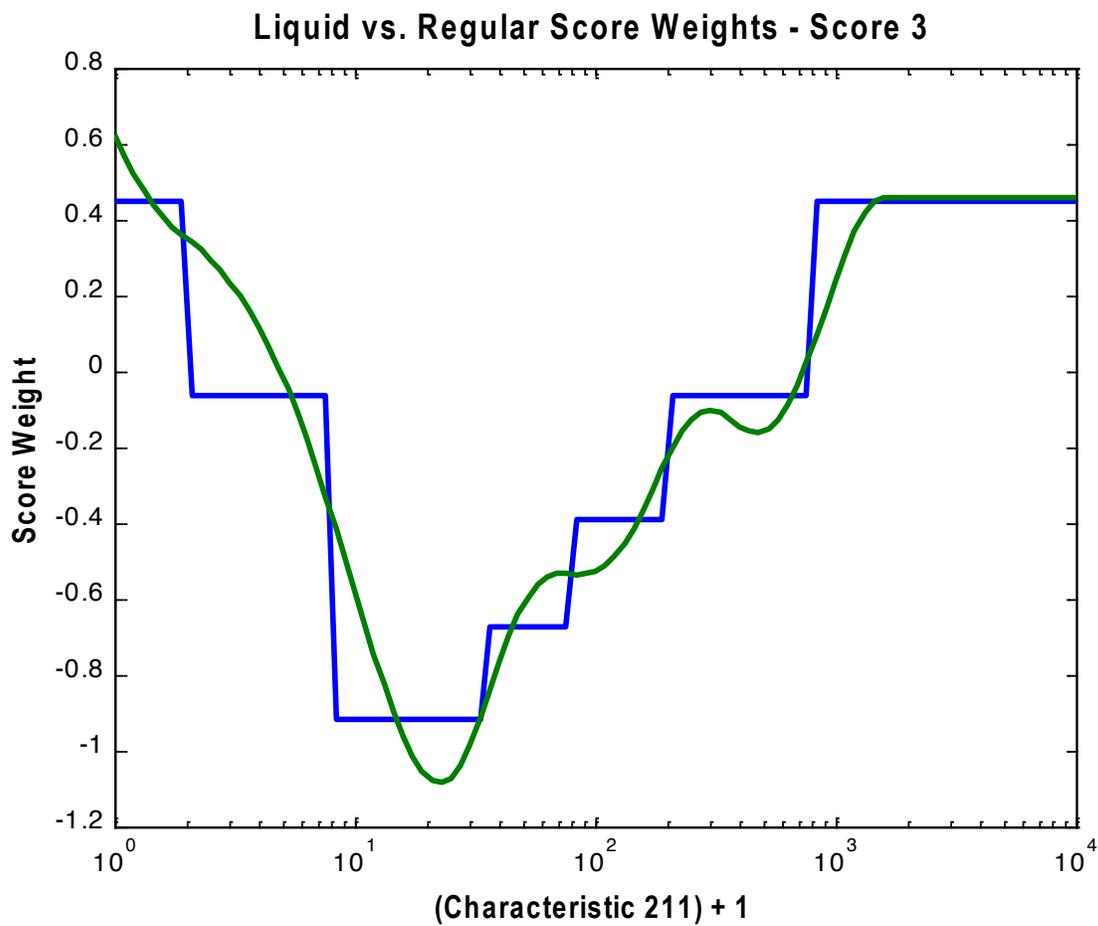

As you can see, the score weight is not monotonically increasing after it hits bottom. To fix this, I changed the pattern constraints to

$$S_{119} \geq ... \geq S_{122}$$
$$S_{124} \leq ... \leq S_{131} \, .$$

The result was

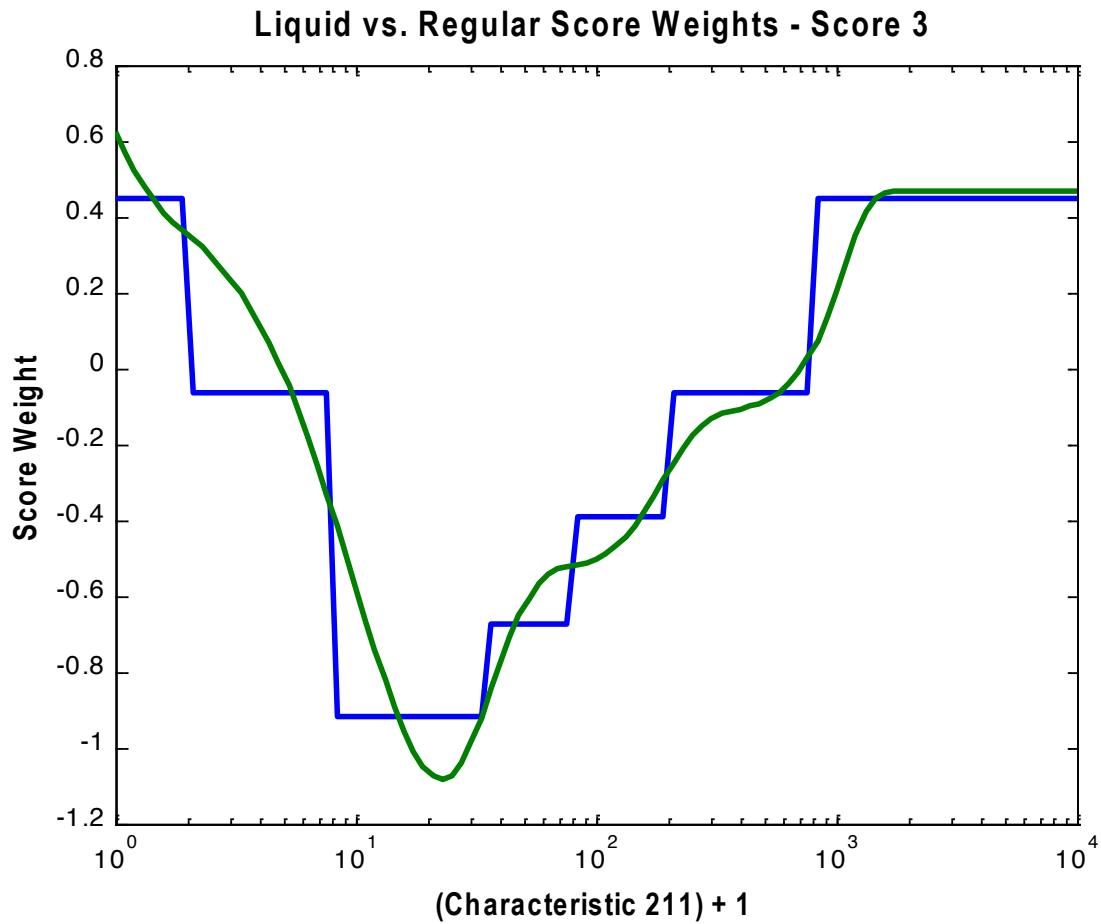

As you can see, the score weight curve is increasing after it hits bottom.

The increasing part of the curve still wiggles a bit. To cure this, we need a penalty term in the objective function, which penalized these kinds of wiggles. This is the subject of future research (see the last Section of this paper).

## 4.10 Initial solution to the quadratic program

The traditional score weight vector, S1, is the solution to problem 1 described in Reference [1]. The values of S1 are shown in Appendix 2 of Reference [1]. Some of the attributes for the liquid scorecard are the same as the attributes in S1. For these, I use the S1 score weights as the initial value for the liquid score coefficients. The other initial values for the liquid score coefficients are set intuitively – taking into account the desired patterns.

The MATLAB code for setting the initial values for S3 (the liquid score coefficients) is

```
        S30=zeros(1,210);
    »   S30(1:25)=S1(65:89);
    »   S30(26:68)=S1(98:140);
    »   S30(69)=S1(1);
    »   S30(70:71)=S1(7:8);
    »   S30(72:73)=S1(14:15);
    »   S30(74:76)=S1(21:23);
    »   S30(77:78)=S1(34:35);
    »   S30(79:80)=S1(42:43);
    »   S30(81)=S1(53);
    »   S30(82)=S1(58);
    »   S30(83)=S1(64);
    »   S30(84)=S1(54);
    »   S30(85)=S1(57);
    »   S30(86:88)=S1(89:91);
    »   S30(89)=S1(97);
    »   S30(90:91)=S1(145:146);
    »   S30(92)=S1(151);
    »   S30(93)=S1(161);
    »   S30(94)=S1(171);
```

```
»  S30(95:210)=[.3 .2 .1 -.1 -.3 -.4 -.6 -.9 ...
                -1.2 -1 -.8 -.6 -.4 -.2 0 .1 ...
                -1.4 -1.2 -1 -.8 -.6 -.4 -.1 .1 ...
          .4 .2 0 -.2 -.4 -.3 -.2 -.1 0 .1 .2 .3 .4
...
                .5 .4 .3 .2 .1 0 -.1 -.2 -.3 ...
          .1 .1 0 0 -.1 -.1 -.2 -.2 -.2 -.3 -.3 -.4
...
                .1 -.1 -.2 -.3 -.4 -.5 -.6 -.8 ...
                -.2 -.2 -.3 -.3 -.4 ...
                .1 0 0 -.1 -.2 -.3 -.4 ...
                -.4 -.3 -.2 -.1 0 0 .1 ...
                -.4 -.3 -.2 -.1 0 .1 .2 ...
                -.1 -.1 0 0 0 0 .1 .1 .1 .2 .2 .2 ...
                -.4 -.3 -.2 -.1 -.1 0 0 .1 .1 .2 .3
.4];
S30=S30';
```

## 4.11 Remaining inputs to the quadratic program

The remaining inputs to the quadratic program are

```
f3=zeros(210,1);
l3=-inf*ones(210,1);
u3=inf*ones(210,1);
b3=zeros(143,1);
beq3=[1.775;zeros(59,1)];
```

## 4.12 Quadratic program

The MATLAB command for running the quadratic program is

```
»  S3=quadprog(H3,f3,A3,b3,Aeq3,beq3,l3,u3,S30);
```

S3 is the liquid score coefficient vector.

## 4.13 Transforming to a WOE scale

The score, S3, is not on a WOE scale. The MATLAB code for putting the score on a WOE scale is

```
»  [beta3 div3 Swoe3]=WOE(S3,Xd,yd);
```

See Appendix 1 for the explanation of the MATLAB function, WOE. Swoe3 is the liquid score coefficient vector on a weight of evidence scale, which is shown in the 7th column of Appendix 2. For the discrete attributes, these liquid score coefficients are very similar to the score weights for the traditional scorecard.

The variable beta3 is the scale factor needed to put the score on a WOE scale and is equal to 1.0006.

### 4.14 Divergence

The above MATLAB code also computed the development divergence, which is 1.7760.

The MATLAB code for computing the validation divergence is

```
»  Divergence(Swoe3,Xv,yv)
```

and the answer is 1.6643.

The validation divergence for the best traditional scorecard was 1.636 (see Score 1 in Appendix 2 of Reference [1]). So the above validation divergence of 1.664 is 1.7% higher. And of course we all know that 1.7% of a billion dollars is a lot of money.

## 5. Plotting the Liquid Parts of the Scorecard

The nice thing about a traditional scorecard is that the score can be represented by a simple list of interpretable score weights for each characteristic. This is not true for the liquid parts of a liquid scorecard. For each liquid characteristic, only the first and last liquid score coefficients are easily interpretable. However, the liquid score can be represented by a simple interpretable graph for each characteristic. In fact, each graph is just a smooth replacement for the step function associated with the traditional score weights.

To analyze the liquid part of the scorecard, I will plot the liquid score weights vs. the traditional score weights for the liquid part of each characteristic. The MATLAB code for developing these plots is described below.

The knots for the liquid part of the liquid characteristics were put into the MATLAB cell array, attk, which is shown in Section 4.5.

The score weights for the traditional scorecard, for each characteristic, are shown in the following MATLAB cell array

```
» atts={S1(2:6) S1(9:13) S1(16:20) S1(24:33) ...
 S1(36:41) S1(44:52) S1(59:63) S1(55:56) ...
S1(93:96) S1(141:144) S1(147:150) S1(152:160) S1(162:170)};
```

The liquid score coefficients for the liquid scorecard, for each characteristic, are shown in the following MATLAB cell array

```
» lc3={Swoe3(95:102) Swoe3(103:110) Swoe3(111:118) ...
      Swoe3(119:131) Swoe3(132:140) Swoe3(141:152) ...
      Swoe3(153:160) Swoe3(161:165) Swoe3(166:172) ...
      Swoe3(173:179) Swoe3(180:186) Swoe3(187:198) ...
      Swoe3(199:210)};
```

For each characteristic there are four plot variables – the x and y-axis for the step function associated with the traditional scorecard, and the x and y-axis for the cubic spline function associated with the liquid scorecard. Each plot variable is represented by a vector of values to be plotted. There are 13 liquid characteristics and therefore 13 plots. The 13 vectors of x's for the 13 step functions will be stored in a cell array called x1. The 13 vectors of y's for the 13 step functions will be stored in a cell array called y1. The 13 vectors of x's for the cubic splines will be stored in a cell array called x2. The 13 vectors of y's for the 13 cubic splines will be stored in a cell array called y2. The MATLAB code for initializing these 4 cell arrays is

```
» x1={0 0 0 0 0 0 0 0 0 0 0 0 0};
» y1={0 0 0 0 0 0 0 0 0 0 0 0 0};
» x2={0 0 0 0 0 0 0 0 0 0 0 0 0};
» y2={0 0 0 0 0 0 0 0 0 0 0 0 0};
```

Note that the initial values in the cell arrays do not have to be vectors.

The MATLAB code for populating these cell arrays is

```
» for i=1:9
    [x1{i} y1{i} x2{i} y2{i}] = ...
     plotvar(1,100,atts{i},attk{i},lc3{i},attk{i});
end
for i=10:13
    [x1{i} y1{i} x2{i} y2{i}] = ...
     plotvar(0,100,atts{i},attk{i},lc3{i},attk{i});
end
```

See Appendix 1 for the details on the MATLAB function, plotvar.

The MATLAB code for plotting the graphs and the resulting graphs are shown below.

```
» plot(x1{1}+1,y1{1},x2{1}+1,y2{1})
```

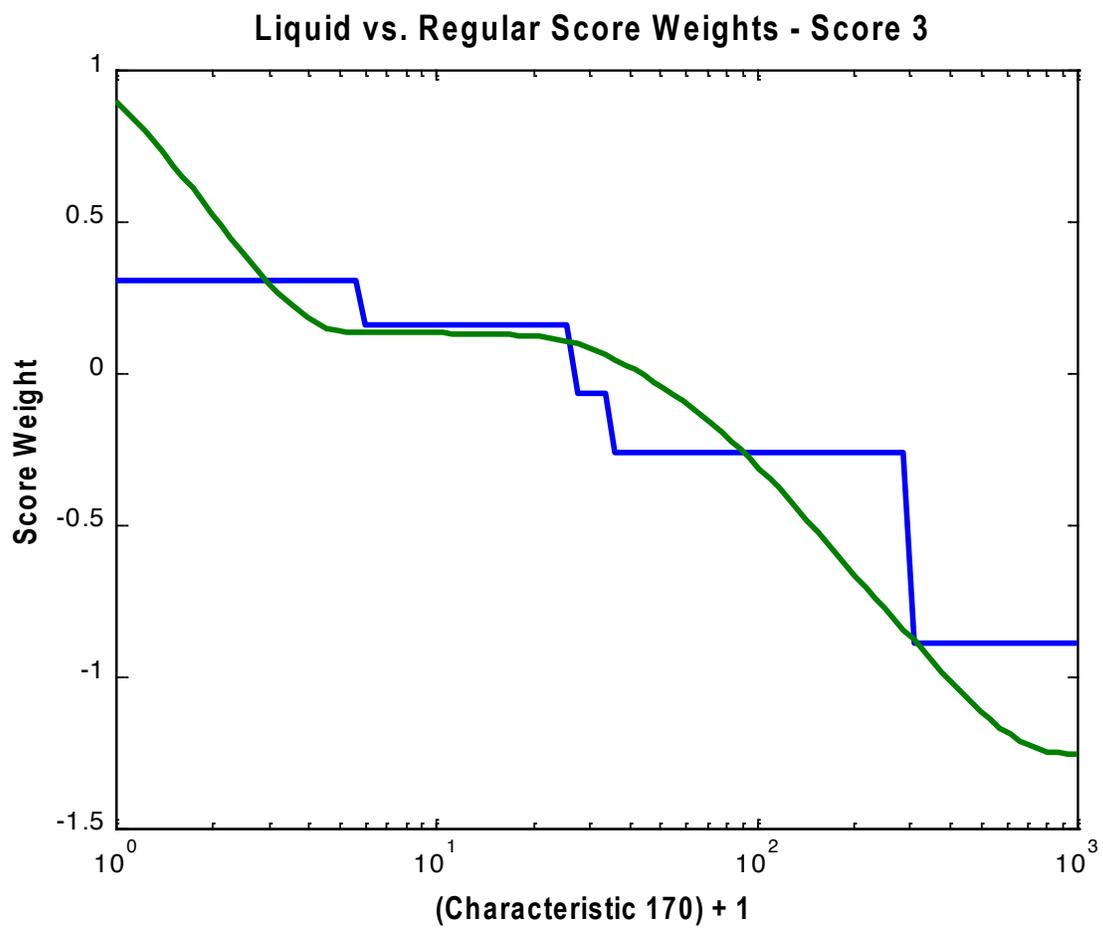

```
» plot(x1{2}+1,y1{2},x2{2}+1,y2{2})
```

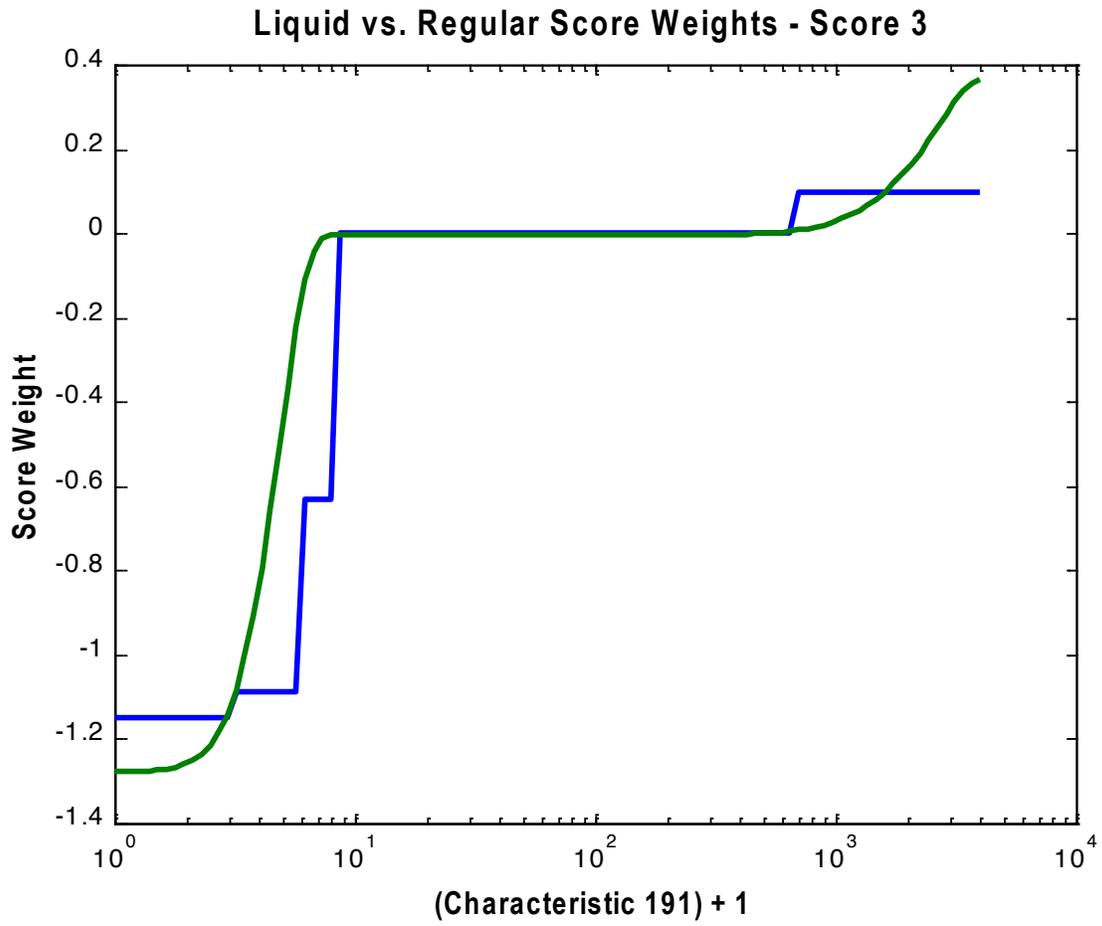

```
» plot(x1{3}+1,y1{3},x2{3}+1,y2{3})
```

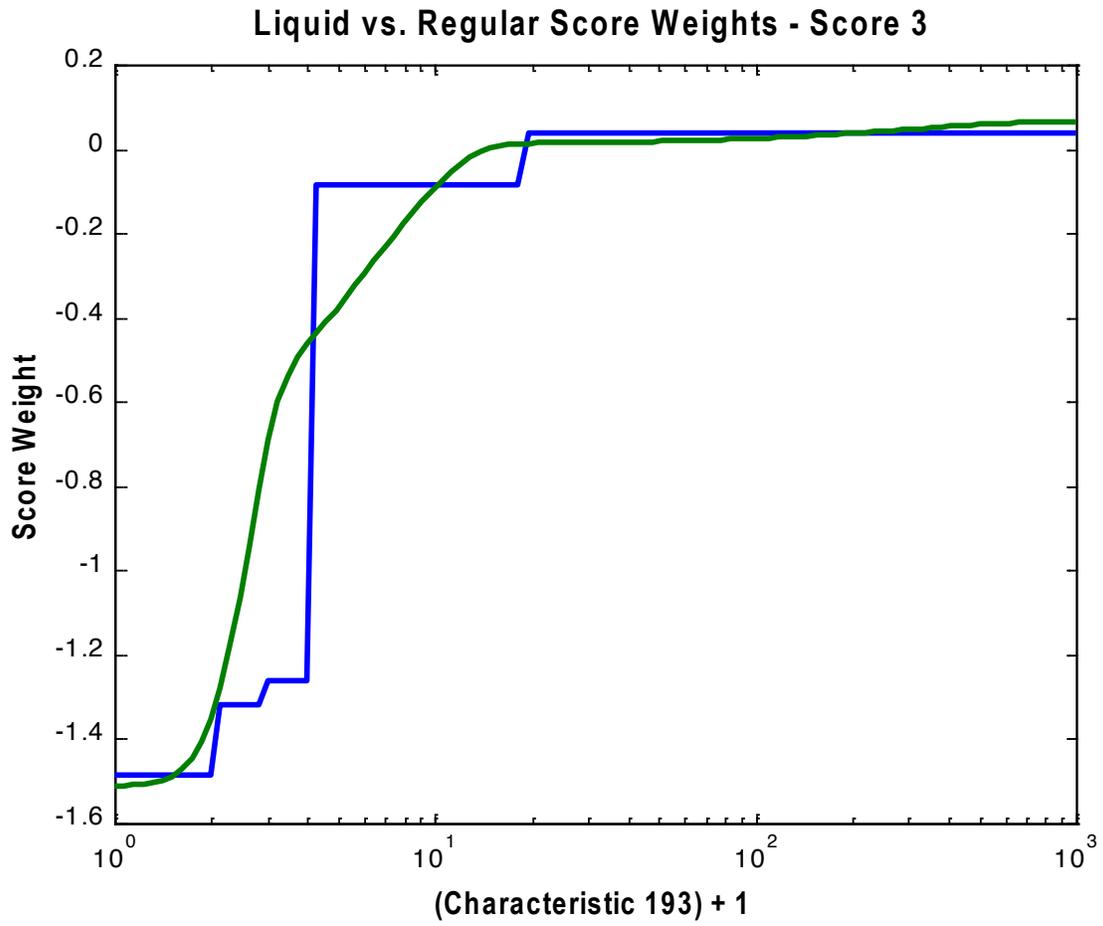

```
» plot(x1{4}+1,y1{4},x2{4}+1,y2{4})
```

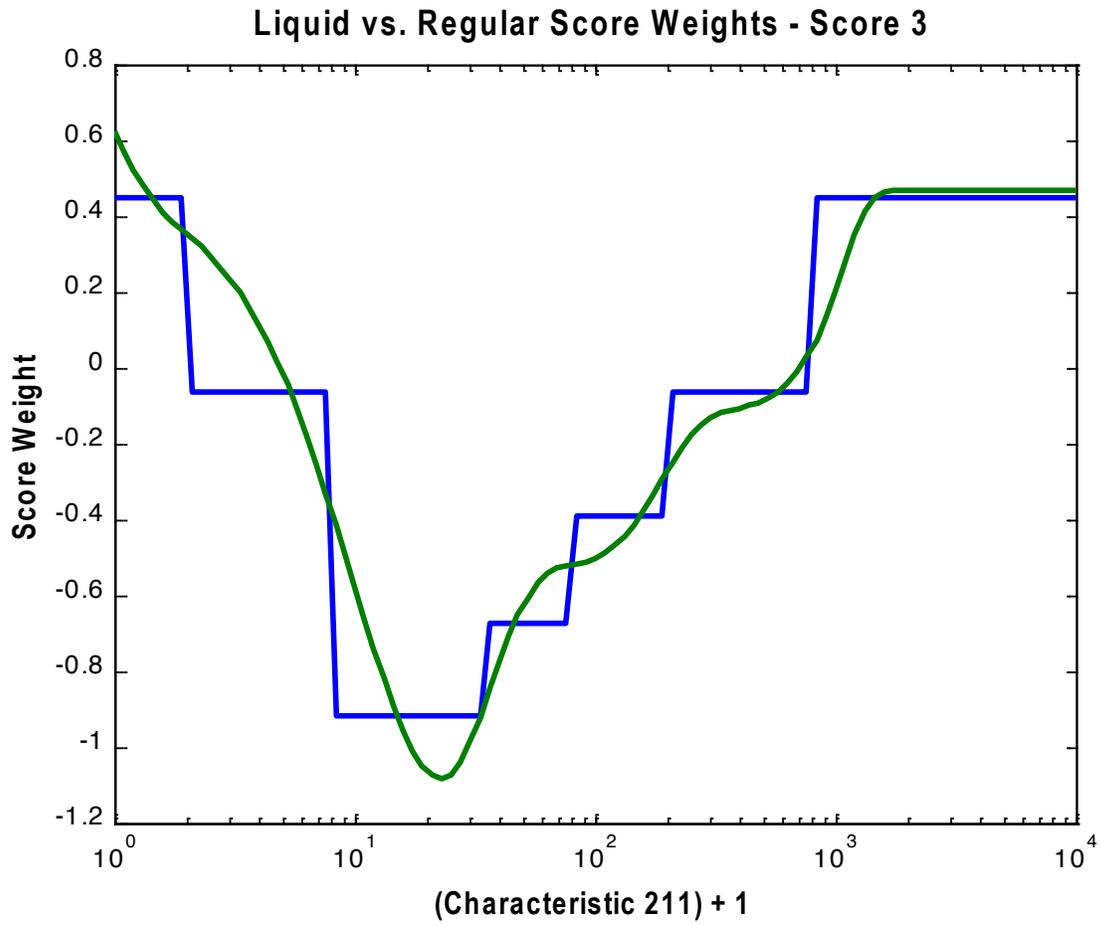

```
» plot(x1{5}+1,y1{5},x2{5}+1,y2{5})
```

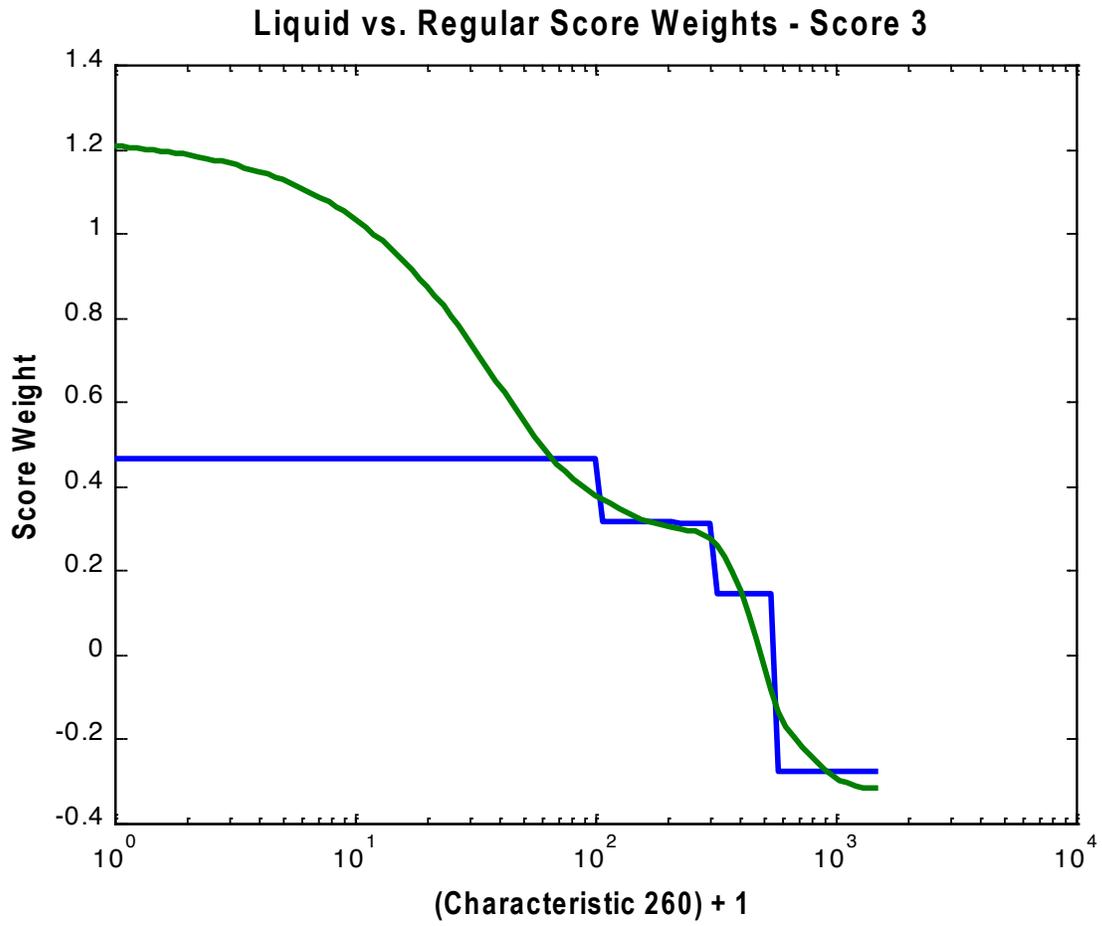

This plot looks very different when the x-axis is plotted on a linear scale.

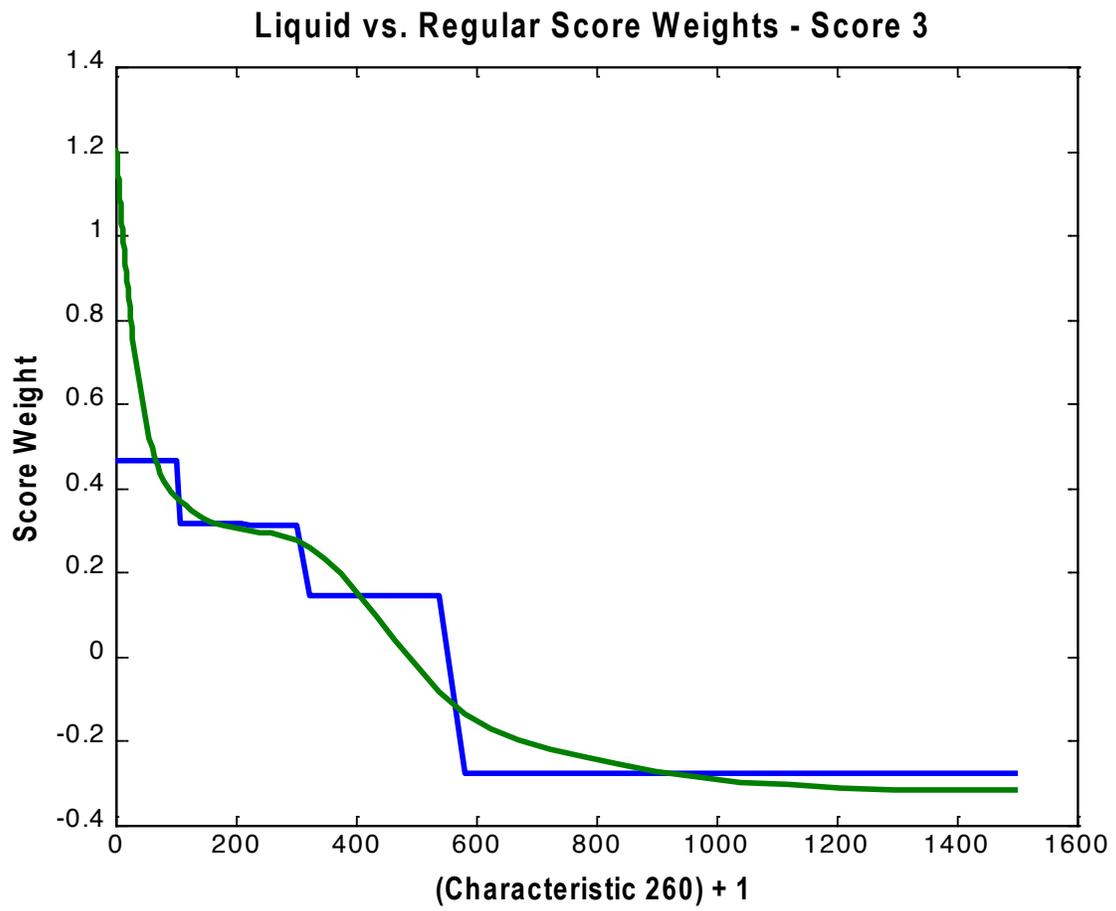

```
» plot(x1{6}+1,y1{3},x2{6}+1,y2{6})
```

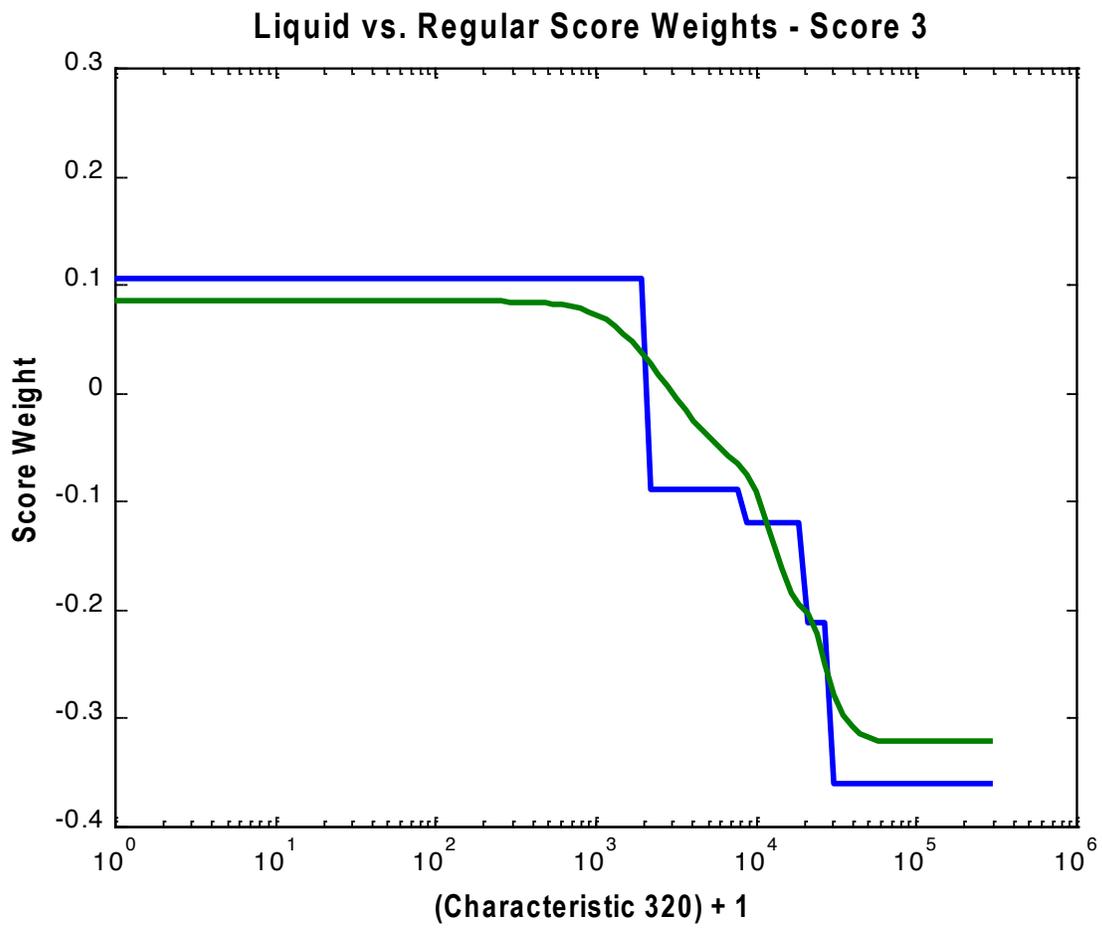

```
» plot(x1{7}+1,y1{7},x2{7}+1,y2{7})
```

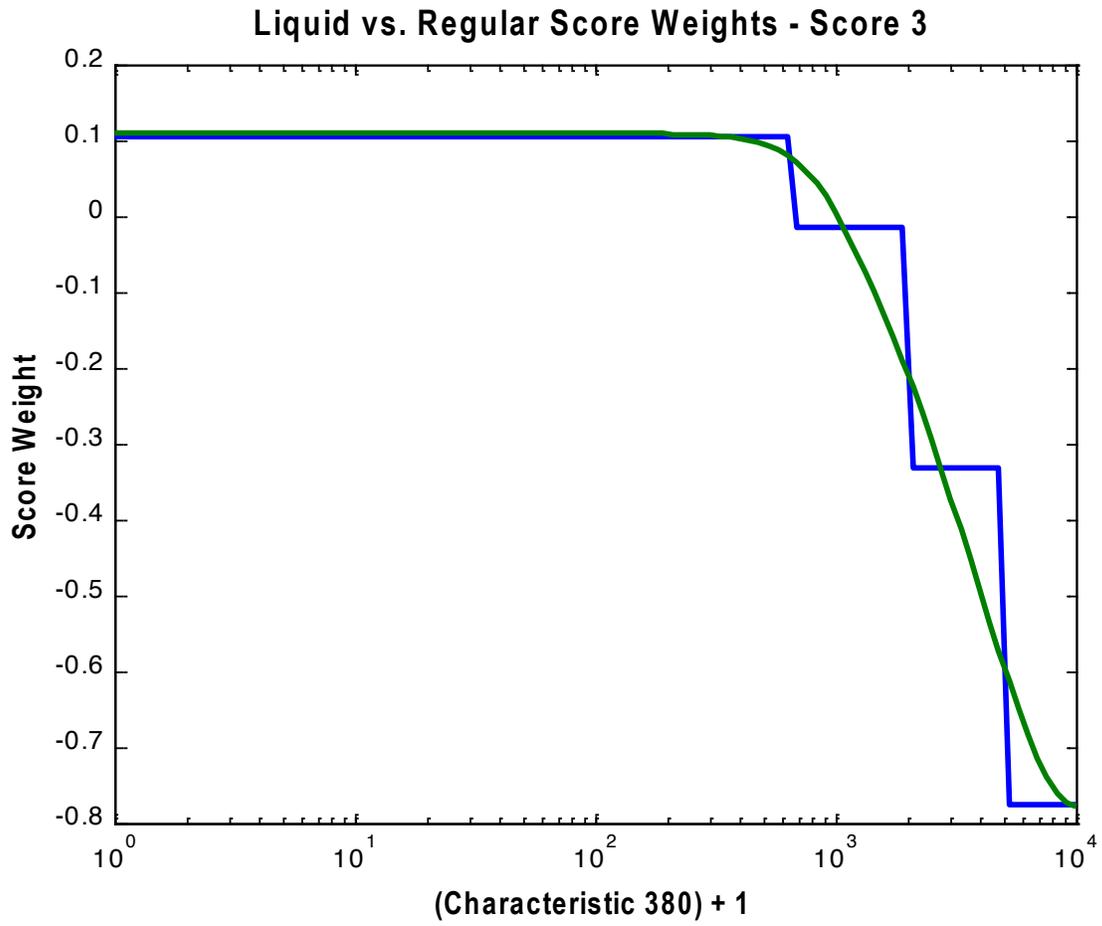

```
» plot(x1{8}+1,y1{8},x2{8}+1,y2{8})
```

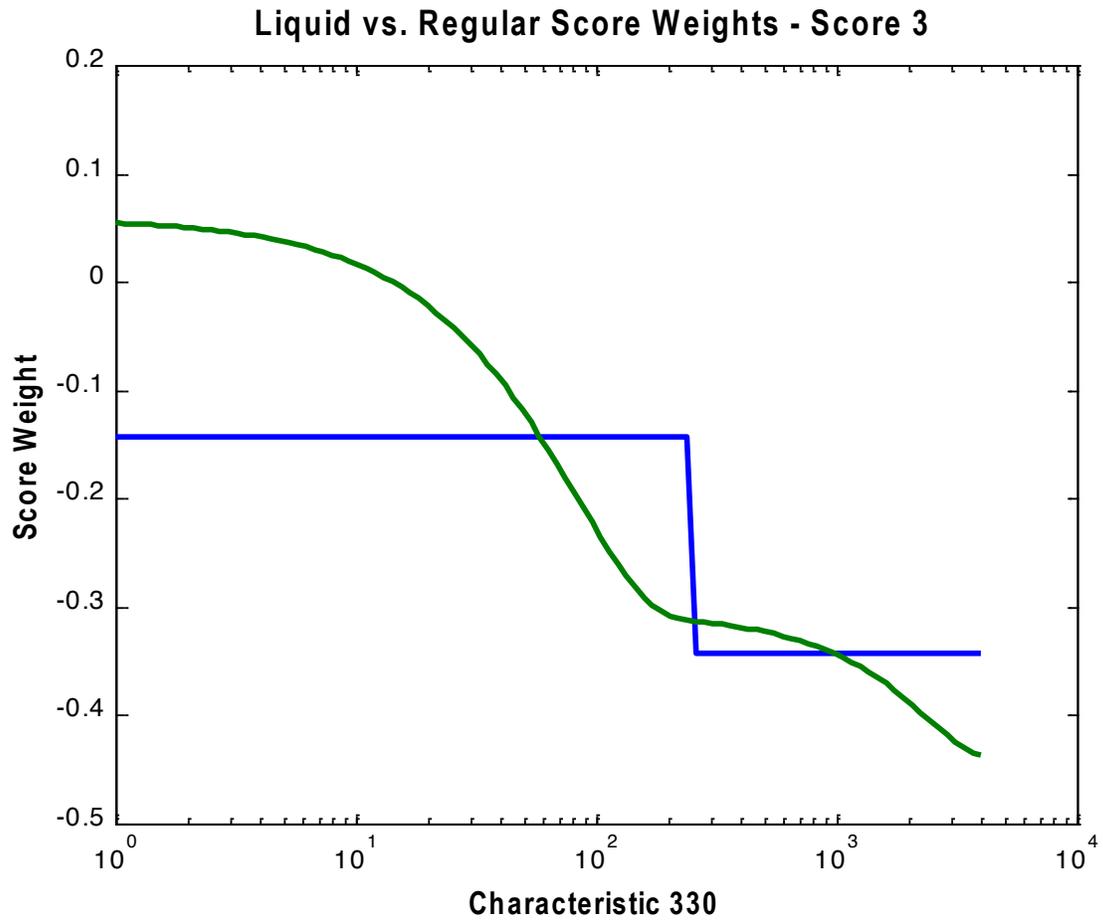

```
» plot(x1{9}+1,y1{9},x2{9}+1,y2{9})
```

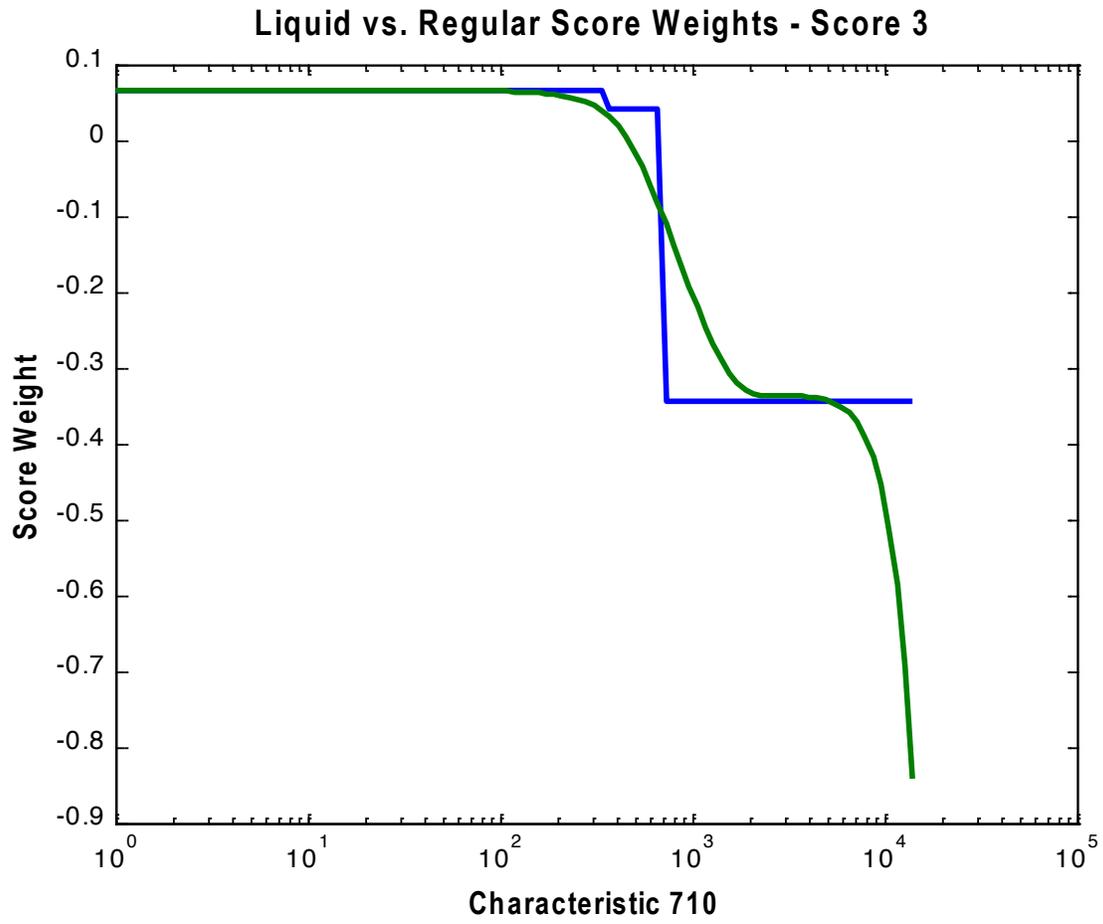

```
» plot(x1{10}+1,y1{10},x2{10}+1,y2{10})
```

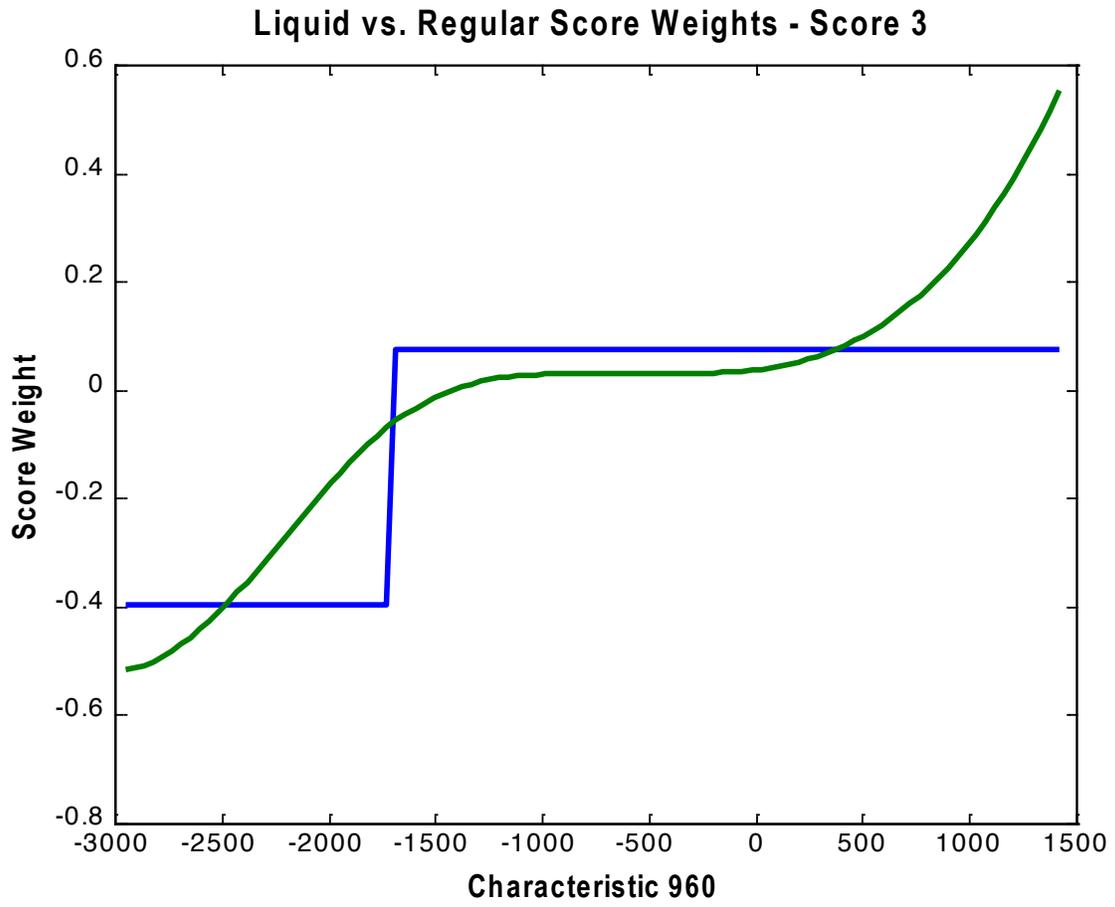

```
» plot(x1{11}+1,y1{11},x2{11}+1,y2{11})
```

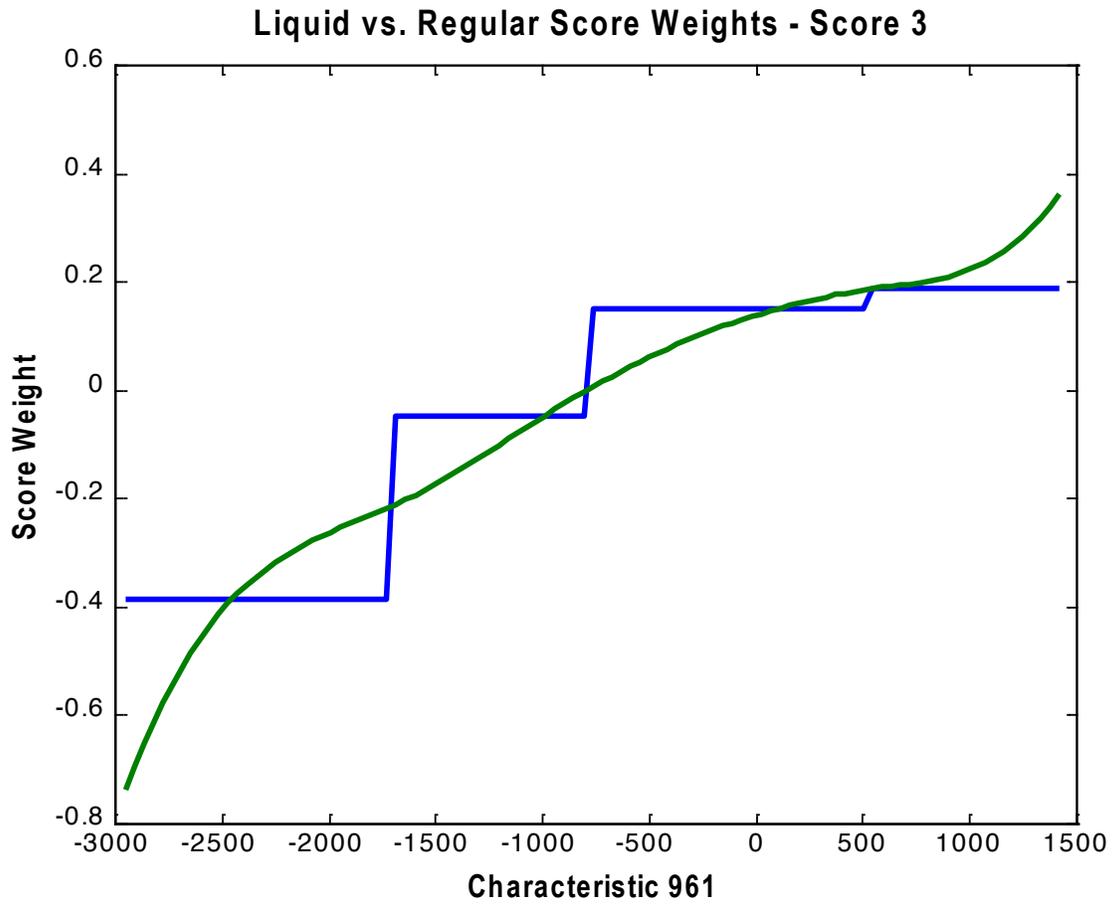

```
» plot(x1{12}+1,y1{12},x2{12}+1,y2{12})
```

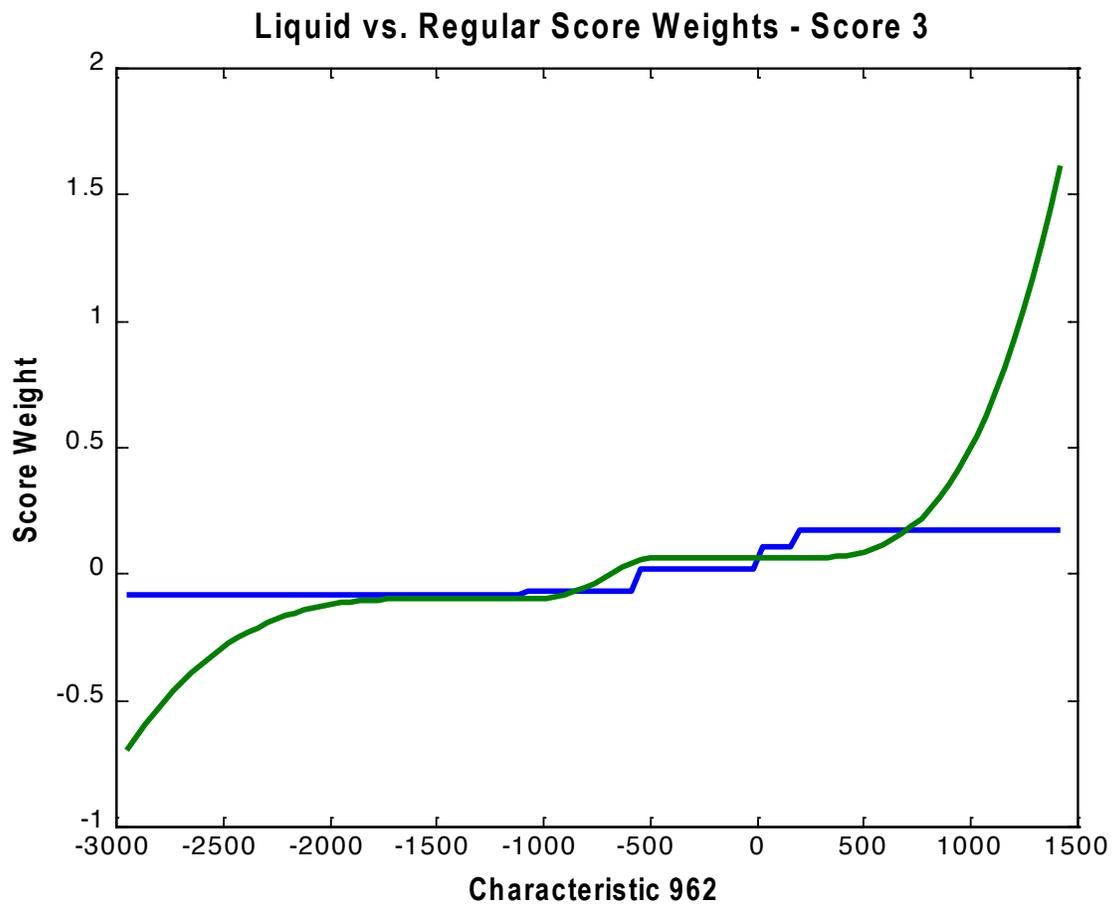

```
» plot(x1{13}+1,y1{13},x2{13}+1,y2{13})
```

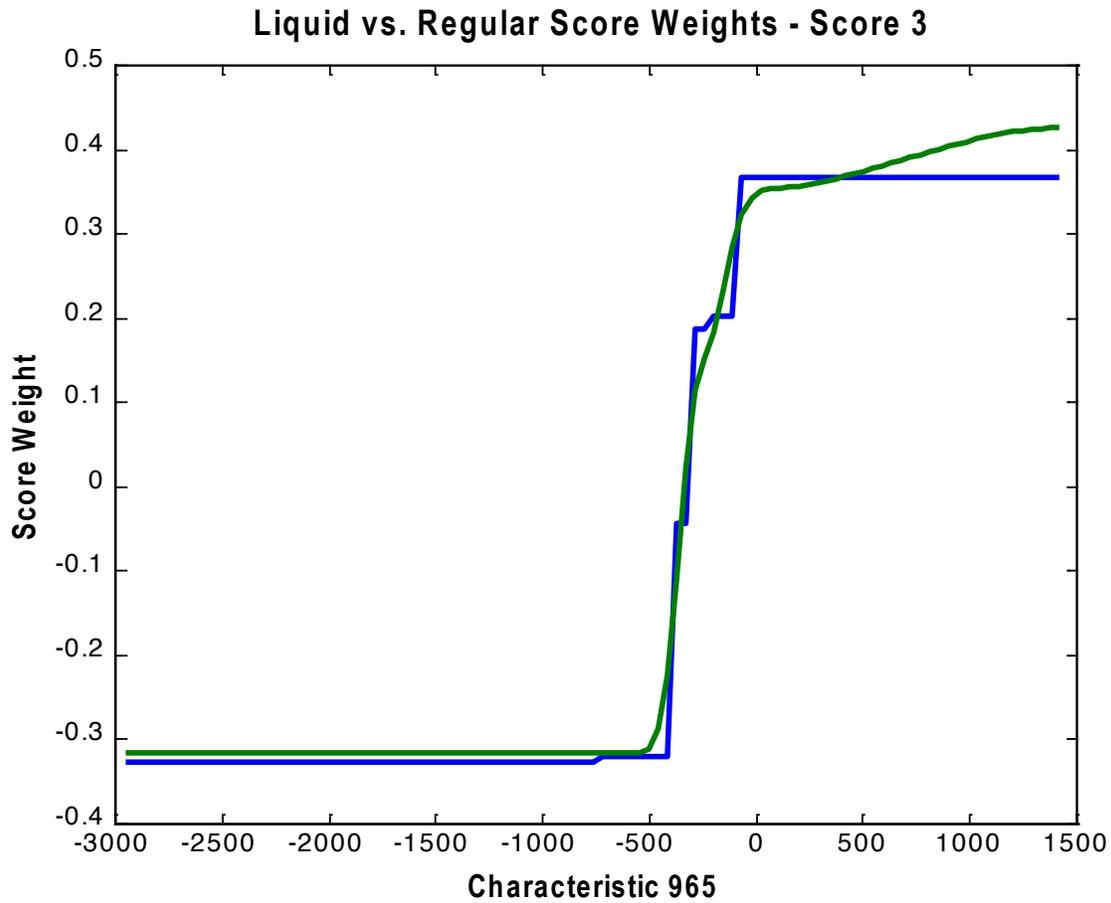

For Characteristics 193, 211, 320, 380, 961, and 965, the cubic splines follow the step functions very closely. For the rest of the characteristics, the cubic splines follow the step functions very closely in the middle of the curve, but deviate from the step functions near one or more of the interval end points.

These deviations may be the source of the extra validation divergence for the liquid scorecard.

# 6. Future Research

## 6.1 Transforming the characteristic scale

In the fraud application used in this paper, many of the plots of the liquid score weights were done on a log scale. However, the cubic spline basis functions were defined on the characteristic scale. It may be a good idea to transform some of the characteristics before defining the B-spline basis functions. For some of the characteristics in the fraud score, I would try $\log(x+1)$ or $\log(x)$ as the transformation. The results could always be transformed back to the original scale for display purposes.

## 6.2 Optimal knots

In this paper, I used the knots of the traditional scorecard to define the cubic spline basis functions. I showed that simple-minded knots, based on percentiles, did not work as well as the scorecard knots. So we know that knot placement is important. The approach taken in Reference [5} may lead to better knots for the cubic spline basis functions.

## 6.3 Smoothness penalty

When fitting curves to data, there should be a natural bias towards smooth curves. For example, the liquid component of Characteristic 211 (plotted in Section 5) is not real smooth on the right side – it wiggles a bit. The notion of, "wiggles too much," means that the derivative of the curve is changing up and down too much; i.e., the magnitude of the second derivative is too high. In classical curve fitting theory, penalizing large second derivatives solves this problem. For the curve, $c(x)$, over the interval, $a \le x \le b$, a traditional measure of second derivative magnitude is

$$\int_a^b \left[c''(x)\right]^2 dx.$$

This can be used in the objective function to penalize large magnitude second derivatives.

In the liquid scorecard application, the curve, $c(x)$, is a linear function of the liquid score coefficients, $s$. So the squared integral measure is a quadratic function of $s$. This fits quite nicely into the quadratic programming score development framework.

However, a little research will be required to figure out the exact formula for this quadratic penalty term. It is very likely that this kind of penalized objective function will lead to better liquid scorecards. Also, it is very likely that this smoothness oriented quadratic penalty term will work better than the simple quadratic penalty described in Reference [1].

## Appendix 1. MATLAB Functions

This appendix contains the MATLAB code for the MATLAB functions used in the liquid scorecard development. Each one has some comments to help in their understanding. I wrote all of these functions.

**bspline**

```matlab
function B=bspline(x,knots)

% Function for computing B-splines at x for knots

n=length(x);
m=length(knots);
L=knots(1);
U=knots(m);

t=[L L L knots U U U];

B=zeros(n,4*(m+2));
v=ones(n,1);

for i=1:(m+1)
    B(:,i)=(t(i)*v<=x)&(x<t(i+1)*v);
end
B(:,m+2)=(t(m+2)*v<=x)&(x<=t(m+3)*v);

for k=2:4
for i=1:(m+2)
    if (t(i+k-1)>t(i))
term1=((x-t(i)*v)/(t(i+k-1)-t(i))).*B(:,i+(m+2)*(k-2));
    else
        term1=zeros(n,1);
    end
```

```
        if (t(i+k)>t(i+1))
  term2=((t(i+k)*v-x)/(t(i+k)-t(i+1))).*B(:,i+1+(m+2)*(k-
2));
        else
            term2=zeros(n,1);
        end

        B(:,i+(m+2)*(k-1))=term1+term2;
    end
    end
```

**bdesign**

```
    function BX=bdesign(x,knots)

    %  x is a nX1 prediction variable.
    %  knots is an array of knots for the cubic ...
       B-spline basis functions.
    %  BX is a matrix of cubic ...
       B-spline basis function predictors.

    m=length(knots);
    t=m+2;
    q=4*t;
    p=3*t+1;

    B=bspline(x,knots);
    BX=B(:,p:q);
```

**divstats**

```
function [C,d,e]=divstats(X,y)

%  X is a n X p design matrix.
%  y is a binary performance variable, ...
   , which is 1{Good}.
%  C is the average covariance matrix ...
   for maximizing divergence.
%  d is used in the divergence constraint.
%  e is used in the centering constraints.

XG=X(y==1,:);
XB=X(y==0,:);

C=(Cov(XG)+Cov(XB))/2;

mG=mean(XG);
mB=mean(XB);
d=mG-mB;
e=mG+mB;
```

**Hmatrix**

```
function H=Hmatrix(C,lambda)

% Computes the H matrix for the quadratic program.
% C is the average covariance matrix.
% lambda is the penalty parameter.
% nsc is the number of liquid score coefficients.

nsc=length(C);

H=2*(C+(lambda/nsc)*eye(nsc));
```

**Aeqiw**

```
function Aeq = Aeqiw(iw,nsparm)

% Computes the Aeq matrix for the in-weights.
% iw is a vector of score parameter indicies
%    used for in weighting.
% nsparm is the number of score parameters.

niw=length(iw);

Aeq=zeros(niw,nsparm);

for i=1:niw
    Aeq(i,iw(i))=1;
end
```

**Aeqcross**

```matlab
function Aeq = Aeqcross(cross,nsparm)

% Computes the Aeq matrix for the cross restrictions.
% cross is a matrix of score parameter indicies,
%    of dimension (ncross) X 2, where row i
%    contains the right and left sides of the
%    ith cross restriction.
% nsparm is the number of score parameters.

ncross=size(cross,1);

Aeq=zeros(ncross,nsparm);

for i=1:ncross
    Aeq(i,cross(i,1))=1;
    Aeq(i,cross(i,2))=-1;
end
```

**Aeqcenter**

```matlab
function Aeq = Aeqcenter(charx,e)

% Computes the Aeq matrix for the centering
constraints.
% charx is a cell array containing the
%   score parameter indicies for each characteristic.
% e is the centering row vector.

nchar=length(charx);
nsparm=length(e);

Aeq=zeros(nchar,nsparm);

for i=1:nchar
    natt=length(charx{i});
    sindex=charx{i};

    for j=1:natt
        Aeq(i,sindex(j))=e(sindex(j));
    end
end
```

**patterns**

```
function Ap=patterns(left,right,ilessthan,nsparm)

%  Computes the A  matrix for the
%    binary pattern constraints,
%    which involve two score parameters.
%  left is a vector of score parameter indicies
%     for the left-hand side of the pattern
constraint.
%  right is a vector of score parameter indicies
%     for the right-hand side of the pattern
constraint.
%  Set it up so that left(i)<right(i).
%  length(right) = length(left).
%  ilessthan is the indicator that pattern i
%     is a < inequality.
%  nsparm is the number of score parameters.

np=length(left);

Ap=zeros(np,nsparm);
coeffl=ones(1,np);

for i=1:np
    coeffl(i)=2*ilessthan(i)-1;
    Ap(i,left(i))=coeffl(i);
    Ap(i,right(i))=-coeffl(i);
End
```

**WOE**

```
function [beta,div,Swoe]=WOE(S,X,y)

%  Function to compute divergence, beta
%    and to transform the score weights
%    to the weight of evidence scale.

%  S is the input score weights.
%  X is a n X p design matrix.
%  y is a binary performance variable,
%      which is 1{Good}.

%  beta is the WOE scale factor.
%  div is the divergence of S and Swoe.
%  Swoe is the WOE score weights.

XG=X(y==1,:);
XB=X(y==0,:);

    scrG=XG*S;
    scrB=XB*S;

    num=mean(scrG)-mean(scrB);
    denom=(cov(scrG)+cov(scrB))/2;

    beta=num/denom;
    div=num*beta;

    Swoe=beta*S;
```

## Divergence

```matlab
function Div=Divergence(S,X,y)

%  Function to compute divergence.

%  S is the input score weights.
%  X is a n X p design matrix.
%  y is a binary performance variable,
%      which is 1{Good}.

XG=X(y==1,:);
XB=X(y==0,:);

    scrG=XG*S;
    scrB=XB*S;

    num=mean(scrG)-mean(scrB);
    denom=(cov(scrG)+cov(scrB))/2;

    beta=num/denom;
    Div=num*beta;
```

**plotvar**

```
function ...
[xsw,sw,xlsw,lsw]=plotvar(logflag,m,attswt,attknots,lcoeff,knots
)

        %  Function for computing plot variables
        %   for comparing regular and liquid score weights.
        %  m+1 is the number of x values.
        %  attswt are the attribute score weights
        %  attknots are the attribute knots.
        %  lcoeff are the liquid score coefficients.
        %  knots are the liquid knots.
        %  xsw is the x-axis for the score weights.
        %  sw are the score weights.
        %  xlsw is the x-axis for the liquid score weights.
        %  lsw are the liquid score weights.

        xsw=xaxis(logflag,m,attknots);
        xlsw=xaxis(logflag,m,knots);

        sw=zeros(1,m+1);
        lsw=zeros(1,m+1);

        for i=1:(m+1)
            sw(i)=stepfcn(xsw(i),attswt,attknots);
            lsw(i)=liqweight(xlsw(i),lcoeff,knots);
        end
```

## xaxis

```
function x=xaxis(logflag,m,knots)

%  Function for computing x axis values
%   spread out appropriately.
%  if logflag=1, then x or x+1 will be plotted on log scale.
%  m+1 is the number of x values.
%  knots give the upper and lower bounds for x.
%  x is a row vector.

nk=length(knots);
zeroflag=0;

if logflag==1
    if knots(1)==0
        zeroflag=1;
        knots=knots+1;
    end
    knots=log10(knots);
end

delta=(knots(nk)-knots(1))/m;
x=knots(1):delta:knots(nk);

if logflag==1
    x=10.^x;
    if zeroflag==1
        x=x-1;
    end
end
```

**stepfcn**

```matlab
function y=stepfcn(x,attswt,attknots)

%  Function for computing the score weight
%    step function
%  attswt are the column vector of
%     attribute score weights.
%  attknots are the row vector of
%     knots for the attributes -
%     including finite lower and upper bounds.
%  length(attswt)=length(attknots)-1.

na=length(attswt);
B=zeros(na,1);

for i=1:(na-1)
    B(i)=(attknots(i)<=x)&(x<attknots(i+1));
end

B(na)=(attknots(na)<=x)&(x<=attknots(na+1));

y=sum(attswt.*B);
```

## liqweight

```
function y=liqweight(x,lcoeff,knots)

%  Function for computing the liquid score weights at
x.
%  lcoeff are the liquid score coefficients.
%  knots are the knots for the B-splines.

B=bspline(x,knots);
y=sum(lcoeff.*B(:,4));
```

# Appendix 2. Structure of the Liquid Scorecard

The following table shows the structure of the liquid scorecard. The traditional attributes for the purely discrete part of the liquid scorecard are in rows 1 through 68. The discrete parts of the liquid characteristics are in rows 69 through 94. And the liquid parts of the liquid characteristics are in rows 95 through 210.

The liquid scorecard was structured in this strange way to facilitate experimentation with the knot structure. However, very little experimentation has been done to date.

## Structure of Liquid Scorecard

| Char | Attribute | Coeff. # | Constraint | Frdata Column | Score Weights for S1 | Score Weights for Swoe3 | | |
|------|-----------|----------|------------|---------------|----------------------|-------------------------|---|---|
| char471 | -9999999 | 1 | = 0 | 100 | 0 | 0 | | |
| char471 | 0 | 2 | < 3 | 101 | -0.429 | -0.4284 | | |
| char471 | 1-<101 | 3 | | 102 | 0.016 | 0.0158 | | |
| char471 | NO INFORMATION | 4 | " = 0 " | 103 | 0 | 0 | | |
| char503 | 0 | 5 | = 26 | 104 | 0.010 | 0.0104 | | |
| char503 | 1-High | 6 | < 5 | 105 | -1.482 | -1.5551 | | |
| char503 | NO INFORMATION | 7 | " = 0 & < 5 " | 106 | 0 | 0 | | |
| char533 | -9999999-<1 | 8 | | 107 | 0.156 | 0.1553 | | |
| char533 | 1 | 9 | > 10 | 108 | -0.359 | -0.3598 | | |
| char533 | 2 | 10 | > 11 | 109 | -0.849 | -0.8283 | | |
| char533 | 3 | 11 | >12 | 110 | -0.909 | -0.9166 | | |
| char533 | 4 | 12 | >13 | 111 | -0.909 | -0.9166 | | |
| char533 | 5-High | 13 | | 112 | -0.909 | -0.9166 | | |
| char533 | NO INFORMATION | 14 | " = 0 " | 113 | 0 | 0 | | |
| char635 | 0 | 15 | | 114 | 0.004 | 0.0065 | | |
| char635 | 1-<3 | 16 | > 17 | 115 | 0.050 | 0.0539 | | |
| char635 | 3 | 17 | > 18 | 116 | 0.050 | 0.0539 | | |
| char635 | 4 | 18 | > 19 | 117 | -0.153 | -0.2009 | | |
| char635 | 5 | 19 | > 20 | 118 | -0.400 | -0.4387 | | |
| char635 | 6-High | 20 | | 119 | -0.619 | -0.7384 | | |
| char635 | NO INFORMATION | 21 | " = 0 " | 120 | 0 | 0 | | |
| char665 | 0 | 22 | > 23 | 121 | 0.106 | 0.1035 | | |
| char665 | 1 | 23 | >24 | 122 | -0.529 | -0.5202 | | |
| char665 | 2-High | 24 | | 123 | -0.572 | -0.5494 | | |
| char665 | NO INFORMATION | 25 | " = 0 " | 124 | 0 | 0 | | |

| | | | | | | | |
|---|---|---|---|---|---|---|---|
| char830 | 0 | 26 | = 46 & >27 | 133 | 0.010 | 0.0104 | |
| char830 | 1 | 27 | > 28 | 134 | 0.009 | 0.0104 | |
| char830 | 2-High | 28 | | 135 | -0.353 | -0.3724 | |
| char830 | NO INFORMATION | 29 | " = 0 " | 136 | 0 | 0 | |
| char835 | 0 | 30 | .>31 | 137 | 0.056 | 0.0598 | |
| char835 | 1 | 31 | >32 | 138 | -0.339 | -0.3746 | |
| char835 | 2 | 32 | > 33 | 139 | -0.420 | -0.4285 | |
| char835 | 3 | 33 | > 34 | 140 | -0.566 | -0.5818 | |
| char835 | 4-High | 34 | | 141 | -0.911 | -0.9558 | |
| char835 | NO INFORMATION | 35 | " = 0 " | 142 | 0 | 0 | |
| char840 | 0 | 36 | > 37 | 143 | 0.235 | 0.2541 | |
| char840 | 1 | 37 | > 38 | 144 | -0.313 | -0.3486 | |
| char840 | 2 | 38 | > 39 | 145 | -1.062 | -1.1408 | |
| char840 | 3-High | 39 | | 146 | -1.497 | -1.5862 | |
| char840 | NO INFORMATION | 40 | " = 0 " | 147 | 0 | 0 | |
| char843 | 1 | 41 | | 148 | 0.737 | 0.7373 | |
| char843 | 2 | 42 | | 149 | -0.188 | -0.1814 | |
| char843 | 3 | 43 | | 150 | 0.001 | -0.0115 | |
| char843 | 4 | 44 | | 151 | -0.013 | -0.0132 | |
| char843 | NO INFORMATION | 45 | " = 0 " | 152 | 0.000 | 0 | |
| char860 | 0 | 46 | = 49 & >47 | 153 | 0.010 | 0.0104 | |
| char860 | 1-High | 47 | | 154 | -0.528 | -0.5538 | |
| char860 | NO INFORMATION | 48 | " = 0 " | 155 | 0 | 0 | |
| char870 | 0 | 49 | > 50 | 156 | 0.010 | 0.0104 | |
| char870 | 1 | 50 | > 51 | 157 | -0.289 | -0.3035 | |
| char870 | 2-High | 51 | | 158 | -0.289 | -0.3035 | |
| char870 | NO INFORMATION | 52 | " = 0 " | 159 | 0 | 0 | |
| char950 | -9999998-<7011 | 53 | > 54 | 160 | 0.659 | 0.6763 | |
| char950 | 3300-<4901 | 54 | > 55 | 161 | 0.324 | 0.3315 | |
| char950 | Travel | 55 | > 56 | 162 | 0.324 | 0.3315 | |
| char950 | 5511-High | 56 | > 57 | 163 | 0.324 | 0.3315 | |
| char950 | MOTO | 57 | > 58 | 164 | -0.007 | 0.0229 | |
| char950 | 5697-<7995 | 58 | > 59 | 165 | -0.079 | -0.0408 | |
| char950 | 3723-<5945 | 59 | > 60 | 166 | -0.079 | -0.0408 | |
| char950 | 5611-<8000 | 60 | > 61 | 167 | -0.079 | -0.0408 | |
| char950 | 4814-<4830 | 61 | > 62 | 168 | -0.079 | -0.0408 | |
| char950 | 5013-<8100 | 62 | > 63 | 169 | -0.157 | -0.1371 | |
| char950 | Gas | 63 | > 64 | 170 | -0.157 | -0.1371 | |
| char950 | 5655-<5949 | 64 | > 65 | 171 | -0.193 | -0.2818 | |
| char950 | 5300-<5942 | 65 | > 66 | 172 | -0.193 | -0.2975 | |
| char950 | 5815-<5963 | 66 | > 67 | 173 | -0.490 | -0.4644 | |
| char950 | 5423-<5655 | 67 | | 174 | -0.734 | -0.6914 | |
| char950 | NO INFORMATION | 68 | " = 0 " | 175 | 0 | 0 | |
| char170 | -9999999 | 69 | " = 0 " | 36 | 0 | 0 | |
| char170 | NO INFORMATION | 70 | " = 0 " | 42 | 0 | 0 | |
| char191 | -9999999 | 71 | " = 0 " | 43 | 0 | 0 | |
| char191 | NO INFORMATION | 72 | " = 0 " | 49 | 0 | 0 | |
| char193 | -9999999 | 73 | | 50 | 0.396 | 0.38 | |
| char193 | NO INFORMATION | 74 | " = 0 " | 56 | 0 | 0 | |
| char211 | -9999999 | 75 | | 57 | -0.096 | -0.1236 | |
| char211 | -9999998 | 76 | | 58 | 0.545 | 0.5488 | |

| char211 | NO INFORMATION | 77 | " = 0 " | 69 | 0 | 0 | |

| | | | | | | | | |
|---|---|---|---|---|---|---|---|---|
| char260 | -9999999 | 78 | | 70 | -0.162 | -0.1628 | | |
| char260 | NO INFORMATION | 79 | " = 0 " | 77 | 0 | 0 | | |
| char320 | -9999999-<0 | 80 | > 141 | 78 | 0.366 | 0.3805 | | |
| char320 | NO INFORMATION | 81 | " = 0 " | 88 | 0 | 0 | | |
| char380 | -9999999-<0 | 82 | = 0 | 93 | 0 | 0 | | |
| char380 | NO INFORMATION | 83 | " = 0 " | 99 | 0 | 0 | | |
| char330 | 0 | 84 | > 161 | 89 | 0.251 | 0.2719 | | |
| char330 | NO INFORMATION | 85 | " = 0 " | 92 | 0 | 0 | | |
| char710 | -9999999 | 86 | " = 0 " | 125 | 0 | 0 | | |
| char710 | -9999998 | 87 | " = 0 " | 126 | 0 | 0 | | |
| char710 | 0 | 88 | > 166 | 127 | 0.066 | 0.0656 | | |
| char710 | NO INFORMATION | 89 | " = 0 " | 132 | 0 | 0 | | |
| char960 | NO INFORMATION | 90 | " = 0 " | 180 | 0 | 0 | | |
| char961 | -9999999 | 91 | | 181 | -0.211 | -0.1416 | | |
| char961 | NO INFORMATION | 92 | " = 0 " | 186 | 0 | 0 | | |
| char962 | NO INFORMATION | 93 | " = 0 " | 196 | 0 | 0 | | |
| char965 | NO INFORMATION | 94 | " = 0 " | 206 | 0 | 0 | | |
| char170 | # | 95 | >96 | | | 0.8988 | | |
| char170 | 0-<5 | 96 | > 97 | | | 0.1314 | | |
| char170 | 5-<25 | 97 | > 98 | | | 0.1314 | | |
| char170 | 25-<35 | 98 | > 99 | | | 0.1314 | | |
| char170 | 35-<300 | 99 | > 100 | | | -0.5095 | | |
| char170 | 300-High | 100 | >101 | | | -1.2554 | | |
| char170 | 1000 | 101 | >102 | | | -1.2554 | | |
| char170 | # | 102 | | | | -1.2554 | | |
| char191 | # | 103 | <104 | | | -1.2762 | | |
| char191 | 0-<2 | 104 | < 105 | | | -1.2762 | | |
| char191 | 2-<5 | 105 | < 106 | | | -1.2762 | | |
| char191 | 5-<7 | 106 | < 107 | | | -0.0049 | | |
| char191 | 7-<650 | 107 | < 108 | | | -0.0049 | | |
| char191 | 650-High | 108 | <109 | | | -0.0049 | | |
| char191 | 4,000 | 109 | <110 | | | 0.3659 | | |
| char191 | # | 110 | | | | 0.3659 | | |
| char193 | # | 111 | <112 | | | -1.5102 | | |
| char193 | 0 | 112 | < 113 | | | -1.5102 | | |
| char193 | 1 | 113 | < 114 | | | -1.5102 | | |
| char193 | 2 | 114 | < 115 | | | -0.5616 | | |
| char193 | 3-<18 | 115 | < 116 | | | 0.0117 | | |
| char193 | 18-High | 116 | <117 | | | 0.0642 | | |
| char193 | 1,000 | 117 | <118 | | | 0.0642 | | |
| char193 | # | 118 | | | | 0.0642 | | |

| char211 | # | 119 | >120 | | | 0.6243 | |
| char211 | 0 | 120 | >121 | | | 0.4267 | |
| char211 | 1-<7 | 121 | >122 | | | 0.1745 | |
| char211 | 7-<35 | 122 | | | | -1.4852 | |
| char211 | 35-<80 | 123 | | | | -0.5049 | |
| char211 | 80-<200 | 124 | | | | -0.5607 | |
| char211 | 200-<400 | 125 | | | | -0.1135 | |
| char211 | 400-<800 | 126 | < 127 | | | -0.1135 | |
| char211 | 800-<1300 | 127 | < 128 | | | 0.0309 | |
| char211 | 1300-<1700 | 128 | < 129 | | | 0.4661 | |
| char211 | 1700-High | 129 | <130 | | | 0.4661 | |
| char211 | 10,000 | 130 | <131 | | | 0.4661 | |
| char211 | # | 131 | | | | 0.4661 | |
| char260 | # | 132 | >133 | | | 1.2064 | |
| char260 | 0-<101 | 133 | > 134 | | | 0.5045 | |
| char260 | 101-<210 | 134 | > 135 | | | 0.338 | |
| char260 | 210-<305 | 135 | > 136 | | | 0.2942 | |
| char260 | 305-<565 | 136 | > 137 | | | 0.2942 | |
| char260 | 565-<700 | 137 | > 138 | | | -0.1331 | |
| char260 | 700-High | 138 | >139 | | | -0.3169 | |
| char260 | 1,500 | 139 | >140 | | | -0.3169 | |
| char260 | # | 140 | | | | -0.3169 | |
| char320 | # | 141 | >142 | | | 0.0843 | |
| char320 | 0-<590 | 142 | > 143 | | | 0.0843 | |
| char320 | 590-<2055 | 143 | > 144 | | | 0.0843 | |
| char320 | 2055-<8405 | 144 | > 145 | | | -0.0502 | |
| char320 | 8405-<16960 | 145 | > 146 | | | -0.0502 | |
| char320 | 16960-<20000 | 146 | > 147 | | | -0.1953 | |
| char320 | 20000-<30000 | 147 | > 148 | | | -0.1953 | |
| char320 | 30000-<40375 | 148 | > 149 | | | -0.2901 | |
| char320 | 40375-<70000 | 149 | > 150 | | | -0.322 | |
| char320 | 70000-High | 150 | >151 | | | -0.322 | |
| char320 | 300,000 | 151 | >152 | | | -0.322 | |
| char320 | # | 152 | | | | -0.322 | |
| char380 | # | 153 | >154 | | | 0.1094 | |
| char380 | 0-<635 | 154 | > 155 | | | 0.1094 | |
| char380 | 635-<1210 | 155 | > 156 | | | 0.1094 | |
| char380 | 1210-<1915 | 156 | > 157 | | | -0.0652 | |
| char380 | 1915-<5000 | 157 | > 158 | | | -0.3542 | |
| char380 | 5000-High | 158 | >159 | | | -0.7033 | |
| char380 | 10,000 | 159 | >160 | | | -0.778 | |
| char380 | # | 160 | | | | -0.778 | |
| char330 | # | 161 | >162 | | | 0.054 | |
| char330 | 1-<250 | 162 | > 163 | | | -0.3068 | |
| char330 | 250-High | 163 | >164 | | | -0.3561 | |
| char330 | 4,000 | 164 | >165 | | | -0.4361 | |
| char330 | # | 165 | | | | -0.4361 | |
| char710 | # | 166 | >167 | | | 0.0656 | |
| char710 | 1-<360 | 167 | > 168 | | | 0.0656 | |
| char710 | 360-<675 | 168 | >169 | | | 0.0656 | |
| char710 | 675-<2435 | 169 | > 170 | | | -0.3371 | |
| char710 | 2435-High | 170 | >171 | | | -0.3371 | |
| char710 | 14,000 | 171 | >172 | | | -0.3371 | |
| char710 | # | 172 | | | | -0.8405 | |
| char960 | # | 173 | <174 | | | -0.5172 | |
| char960 | -2,950 -1700 | 174 | < 175 | | | -0.4957 | |
| char960 | -1700-<-800 | 175 | < 176 | | | 0.0294 | |
| char960 | -800-<-450 | 176 | < 177 | | | 0.0294 | |
| char960 | " -450-<High " | 177 | <178 | | | 0.0294 | |
| char960 | 1,425 | 178 | <179 | | | 0.0294 | |
| char960 | # | 179 | | | | 0.552 | |
| char961 | # | 180 | <181 | | | -0.74 | |
| char961 | -2,950-<-1700 | 181 | < 182 | | | -0.3012 | |
| char961 | -1700-<-800 | 182 | < 183 | | | -0.2821 | |
| char961 | -800-<550 | 183 | < 184 | | | 0.0584 | |
| char961 | 550-High | 184 | <185 | | | 0.2021 | |
| char961 | 1,425 | 185 | <186 | | | 0.2021 | |
| char961 | # | 186 | | | | 0.3608 | |
| char962 | # | 187 | <188 | | | -0.7003 | |
| char962 | -2,950 -1500 | 188 | < 189 | | | -0.0999 | |
| char962 | -1500-<-1100 | 189 | < 190 | | | -0.0999 | |
| char962 | -1100-<-850 | 190 | < 191 | | | -0.0999 | |
| char962 | -850-<-550 | 191 | < 192 | | | -0.0999 | |
| char962 | -550-<-400 | 192 | < 193 | | | 0.062 | |
| char962 | -400-<-300 | 193 | < 194 | | | 0.062 | |
| char962 | -300-<1 | 194 | < 195 | | | 0.062 | |
| char962 | 1-<200 | 195 | < 196 | | | 0.062 | |
| char962 | 200-High | 196 | <197 | | | 0.062 | |
| char962 | 1,425 | 197 | <198 | | | 0.062 | |
| char962 | # | 198 | | | | 1.6151 | |
| char965 | # | 199 | <200 | | | -0.3167 | |
| char965 | -2,950 -950 | 200 | < 201 | | | -0.3167 | |
| char965 | -950-<-750 | 201 | < 202 | | | -0.3167 | |
| char965 | -750-<-550 | 202 | < 203 | | | -0.3167 | |
| char965 | -550-<-400 | 203 | < 204 | | | -0.3167 | |
| char965 | -400-<-300 | 204 | < 205 | | | -0.3167 | |
| char965 | -300-<-200 | 205 | < 206 | | | 0.149 | |
| char965 | -200-<-100 | 206 | < 207 | | | 0.149 | |
| char965 | -100-<80 | 207 | < 208 | | | 0.3518 | |
| char965 | 80-High | 208 | <209 | | | 0.3518 | |
| char965 | 1,425 | 209 | <210 | | | 0.4246 | |
| char965 | # | 210 | | | | 0.4246 | |
| | | | | | | | |
| | | | | | | | |
| | | | | | | | |